\documentclass[12pt]{article}

\usepackage{amsmath,amssymb,theorem,epsfig}
\usepackage{amscd}

\numberwithin{equation}{section}

\theorembodyfont{\rmfamily}

\newcommand{\sfrac}[2]{{\textstyle\frac{#1}{#2}}}
\newcommand{\ns}{\normalsize}

\newcommand{\hf}{\frac{1}{2}}
\newcommand{\shf}{\sfrac{1}{2}}

\newcommand{\ZZ}{{\mathbb{Z}}}
\newcommand{\PP}{{\mathbb{CP}}}

\newcommand{\id}{{\text{id}}}
\newcommand{\ch}{{\text{ch}}}

\newcommand{\cE}{{\mathcal{E}}}
\newcommand{\cN}{{\mathcal{N}}}
\newcommand{\cS}{{\mathcal{S}}}
\newcommand{\cL}{{\mathcal{L}}}

\newcommand{\tX}{{\widetilde{X}}}
\newcommand{\tV}{{\widetilde{V}}}

\newcommand{\sm}{
   {SU(3)_{\mathrm{C}}{\times} {SU(2)_{\mathrm{L}}}{\times} {U(1)_{\mathrm{Y}}}}}
\newcommand{\fixedset}{\mathcal{F}_{\tau_B}}

\def\a{\alpha}
\def\b{\beta}
\def\g{\gamma}

\def\d{\delta}

\def\z{\psi}
\def\k{\kappa}
\def\l{\lambda}

\def\o{\omega}
\def\p{\pi}

\def\s{\sigma}
\def\t{\tau}

\def\z{\zeta}

\def\D{\Delta}

\def\G{\Gamma}

\begin{document}


\begin{titlepage}

\title{
   \vspace*{-5em}   
   \vfill
   {\LARGE Standard Models from Heterotic M-theory}
       \\ } 
\author
   {\ns Ron Donagi$^1$, Burt A.~Ovrut$^2$, Tony Pantev$^1$ 
      and Daniel Waldram$^{3\;4}$ \\[0.5em]
   {\it\ns $^1$Department of Mathematics, 
      University of Pennsylvania} \\[-0.7em]
      {\it\ns Philadelphia, PA 19104--6395, USA}\\[-0.3em]
   {\it\ns $^2$Department of Physics, 
      University of Pennsylvania} \\[-0.7em]
      {\it\ns Philadelphia, PA 19104--6396, USA}\\[-0.3em]
   {\it\ns $^3$Department of Physics, 
      Joseph Henry Laboratories}\\[-0.7em]
      {\it\ns Princeton University, Princeton, NJ 08540, USA}\\[-0.3em]
   {\it\ns $^4$Theory Division, CERN CH-1211, Geveva 23, Switzerland}
      \footnote{current address}}

\date{}

\maketitle

\begin{abstract}
We present a class of $N=1$ supersymmetric models of particle physics,
derived directly from heterotic M-theory, that contain three families
of chiral quarks and leptons coupled to the gauge group $\sm$. These
models are a fundamental form of ``brane-world'' theories, with an
observable and hidden sector each confined, after compactification on
a Calabi--Yau threefold, to a BPS threebrane separated by a
five-dimensional bulk space with size of the order of the intermediate 
scale. The requirement of three families, coupled to the fundamental
conditions of anomaly freedom and supersymmetry, constrains these
models to contain additional fivebranes wrapped around holomorphic
curves in the Calabi--Yau threefold. These fivebranes ``live'' in the
bulk space and represent new, non-perturbative aspects of these
particle physics vacua. We discuss, in detail, the relevant
mathematical structure of a class of torus-fibered Calabi--Yau
threefolds with non-trivial first homotopy groups and construct
holomorphic vector bundles over such threefolds, which, by including
Wilson lines, break the gauge symmetry to the standard model gauge group.
Rules for constructing phenomenological particle physics models in
this context are presented and we give a number of explicit examples.
\end{abstract}

\noindent CERN-TH/99-407, UPR-853T \hfill

\noindent December 1999 \hfill

\thispagestyle{empty}

\end{titlepage}


\section{Introduction}

In fundamental work, it was shown by Ho\v rava and Witten
\cite{HW1,HW2} that if eleven-dimensional M-theory is compactified on
the orbifold $S_1/Z_2$, a chiral ${\cal{N}}=1$, $E_{8}$ gauge supermultiplet
must exist in the twisted sector of each of the two ten-dimensional
orbifold fixed planes. They argued that this gave the low-energy
description of the strongly coupled $E_8\times E_8$ heterotic
string. This is relevant for phenomenological particle physics since,
when compactified to four dimensions~\cite{W}, such theories will
exhibit the left-right asymmetry of quark and lepton electroweak couplings
required to describe the weak interactions. It is important to note
that, in this theory, the chiral gauge matter is confined solely to
the orbifold planes, while pure supergravity inhabits the bulk space
between these planes. Thus, Ho\v rava-Witten theory is a concrete and
fundamental representation of the idea of a ``brane-world''. 

Witten then showed~\cite{W} that, if further compactified to four
dimensions on a Calabi--Yau threefold, the ${\cal{N}}=1$ supersymmetric low
energy theory exhibits realistic gauge unification and gravitational
coupling strength provided the Calabi--Yau radius, $R$, is of the
order of inverse $10^{16}$GeV and that the orbifold radius, $\rho$, is larger
than $R$. Estimates of $\rho/R$ vary from 4-5 to several orders of
magnitude~\cite{W,scales}. Thus, Ho\v rava--Witten theory has a ``large''
internal bulk dimension, although it is of order the intermediate
scale and not the inverse TeV-size, or larger, bulk dimensions,
discussed recently~\cite{dim,RS}.

As in the case of the weakly coupled heterotic string, when
compactifying the Ho\v rava--Witten theory to lower dimensions, 
it is possible that all
or, more typically, a subset of the $E_8$ gauge fields do not vanish
classically in the internal Calabi--Yau threefold directions. Since
these gauge fields ``live'' on the Calabi--Yau 
manifold, $3+1$-dimensional Lorentz invariance is left
unbroken. Furthermore, demanding that the associated field
strengths satisfy the ``Hermitian Yang--Mills'' constraints
$F_{ab}=F_{\bar{a}\bar{b}}=g^{a\bar{b}}F_{a\bar{b}}=0$, ensures that
$\cN=1$ supersymmetry is preserved. These gauge field vacua have
two important effects. First, they spontaneously break the $E_8$ gauge
group. Suppose that the non-vanishing gauge fields are associated with
the generators of a subgroup $G\subseteq E_8$. Then, at low energies,
the $E_8$ gauge group is spontaneously broken to the commutant
subgroup $H$, formed from those generators of $E_8$ which commute with
$G$. This allows one, in principle, to reduce the $E_8$ gauge group to
smaller and phenomenologically more interesting gauge groups such as
unification groups $E_6$, $SO(10)$ and $SU(5)$ as well as the
standard model gauge group $\sm$. The second effect is on the spectrum
of massless charged particles in the low-energy theory which arise from
the dimensional reduction of fields associated with the broken
generators. By choosing the gauge vacuum, one can control the number
of families of chiral quarks and leptons on the orbifold fixed plane. 

A new ingredient in compactifications of Ho\v rava--Witten theory is
that there may also be M-theory fivebranes in the
vacuum~\cite{W,nse}. Requiring that supersymmetry and the
$3+1$-dimensional Lorentz invariance is unbroken restricts the
form of the fivebranes. Their six-dimensional worldvolumes are
required to span the external $3+1$-dimensional space while wrapping around a
real two-dimensional surface $W$ within the Calabi--Yau manifold. Thus each
fivebrane is at a definite point in the $S^1/Z_2$ orbifold. Furthermore,
in order to preserve ${\cal{N}}=1$ supersymmetry,
the real surface $W$ in the Calabi--Yau threefold must be
holomorphic~\cite{W}, meaning it is defined by the vanishing of
functions which depend only on the holomorphic and not the
anti-holomorphic coordinates of the Calabi--Yau space. Since the
surface has one complex dimension, we will often refer to it as a
(complex) curve. 

There is an important cohomological constraint which relates the
number of fivebranes and the curves on which they are wrapped to
properties of the gauge vacuum and the Calabi--Yau
manifold~\cite{W}. Without fivebranes, this becomes a constraint
directly relating the gauge vacuum to the Calabi--Yau space. One of
the important properties of including fivebranes in the background is
that it relaxes this condition, making it much easier to find gauge
vacua with suitable unbroken gauge groups and low-energy particle
content. As we will see, all the examples of realistic
compactifications we will give necessarily include fivebranes.

We refer to the general compactification of Ho\v rava--Witten
theory on backgrounds with arbitrary supersymmetric gauge fields
and M~fivebranes as ``heterotic M-theory''. The simplest heterotic 
M-theory vacuum, with the spontaneous breaking of $E_{8}$ to
$E_{6}$ by taking $G=SU(3)$ and identifying it with the spin
connection of the Calabi--Yau threefold, the ``standard
embedding'', was discussed in the original paper of
Witten~\cite{W}. Such vacua do not contain fivebranes. A
discussion of general gauge vacua in this context and their
low-energy implications was presented in~\cite{nse,lpt}, with some
explicit orbifold examples given in~\cite{stei}. Fivebranes were first
introduced into heterotic M-theory vacua, and their properties
discussed, in~\cite{nse}.  

An important result of this analysis is that there is an effective
five-dimensional theory~\cite{losw1,losw2} which realizes the
heterotic M-theory vacua as BPS ``brane-worlds''. The bulk theory is gauged
supergravity with a negative cosmological constant, and the orbifold
planes and fivebranes appear as an array of BPS three-branes domain
walls. The branes coming from the orbifold planes can have negative
tension and the full spacetime solution is very similar to the anti-de
Sitter space solution of Randall and Sundrum~\cite{RS}. In fact, if
one chooses to fix the Calabi--Yau moduli, it precisely realizes the
solution of~\cite{RS}.  

Having defined heterotic M-theory, the real question is whether
explicit heterotic M-theory vacua exist with the required low-energy
properties. As for weakly coupled heterotic string compactifications,
the difficult 
part in finding such models is in constructing the gauge vacuum. Given
a Calabi--Yau threefold $Z$, what supersymmetric non-Abelian gauge
field vacuum configurations associated with a subgroup $G\subseteq
E_{8}$ can be defined on it? Unfortunately, solving the 
six-dimensional Hermitian Yang--Mills
constraints explicitly is rarely possible, even in flat space. 
One, therefore, must look for an alternative
construction of these Yang-Mills connections. Such an alternative
is to be found in the work of Donaldson \cite{Don} and Uhlenbeck and Yau
\cite{UhYau}, which recasts the problem in terms of holomorphic vector
bundles. These authors proved that for each holomorphic vector bundle,
with structure group $G$ over $Z$, satisfying the condition of being
``semi-stable'', there exists a solution to the six-dimensional
Hermitian Yang--Mills equations, and conversely. Thus, the analytic
problem of finding gauge vacua by solving the Hermitian Yang--Mills
constraint equations over $Z$ is reduced to an essentially 
topological problem of constructing semi--stable holomorphic vector 
bundles over the same manifold. 

It is not immediately clear that any simplification has been achieved,
but indeed it has, since some methods for constructing semi-stable
holomorphic vector bundles are known. In particular, bundles were 
constructed for Calabi--Yau manifolds which appear as complete
intersections in weighted projective
spaces~\cite{dg,kachru}. Recently, however, there were important new
results by several authors~\cite{FMW,D,BJPS} on the construction of
semi-stable holomorphic vector bundles over elliptically fibered
Calabi--Yau manifolds. These are manifolds with a fibered structure,
where the fiber is a two-dimensional torus, and which admit a global
section. They are of particular interest because the heterotic string
compactified on such spaces has a dual F-theory description. One
method, the spectral-cover construction, essentially uses T-duality on
the elliptic fiber to form the bundle from a simpler T-dual
configuration. The simplest examples are for structure groups
$SU(n)\subset E_8$, but other structure subgroups are possible as
well. Thus, using holomorphic vector bundles and the
Donaldson--Uhlenbeck--Yau theorem, it has been possible to classify
and give the properties of a large class of $SU(n)$ gauge vacua even
though the associated solutions of the Yang--Mills equations are
unknown. This one might call the ``mathematicians'' approach, but, at
present, it seems to be by far the simplest solution to an important
physical problem. 

The new results on bundles on elliptically fibered Calabi--Yau
threefolds allows one to construct a number of phenomenologically
interesting gauge vacua. In~\cite{don1,don2}, using the construction
of~\cite{FMW,D,BJPS} and results of~\cite{curio,ba},
three-family vacua with unification groups such as $E_{6}$, $SO(10)$
and $SU(5)$ were obtained, corresponding to vector bundle structure
groups $SU(3)$, $SU(4)$ and $SU(5)$ respectively. The classical moduli
space of the fivebranes in such vacua was then discussed
in~\cite{don3}. However, it was not possible to break $E_{8}$ directly
to the standard gauge group $\sm$ in this manner.

A natural solution to this problem, and that utilized in this paper, is to
use non-trivial Wilson lines to break the GUT group down to the
standard gauge group \cite{bos,W85}. This requires that the fundamental
group of the Calabi--Yau threefold be non-trivial. 
Unfortunately, one
can show that all elliptically fibered Calabi--Yau threefolds are
simply connected, with the exception of fibrations over an
Enriques base. In this case, however, we demonstrated in~\cite{don2}
that the vacuum obtained via spectral covers is not consistent 
with the requirement of three
families of quarks and leptons. Further progress clearly required
resolution of this fundamental problem.

With this in mind, recall that an elliptic fibration is simply a torus
fibration that admits a zero section. As we noted above, the
requirement of a zero section severely restricts the fundamental group
of the threefold to be, modulo the Enriques exception, trivial. However, 
if one lifts the zero-section requirement, and considers holomorphic
vector bundles over torus-fibered Calabi--Yau threefolds without
section, then one expects to find non-trivial first homotopy groups
and Wilson lines in vacua that are consistent with the three-family
requirement. 

In this paper, we give the relevant mathematical properties of
a specific class of torus-fibered Calabi--Yau threefolds without
section and construct holomorphic vector bundles over such
threefolds. The technique is familiar from standard heterotic string
constructions~\cite{W85}. One starts with an elliptically fibered
Calabi--Yau threefold $X$ which has trivial fundamental group, but is
chosen to have a discrete group of freely acting symmetries
$\G$. Modding out, by identifying points on $X$ related by $\G$, one
forms a new, smooth threefold $Z=X/\G$ which has fundamental group
$\p_1(Z)=\G$. However, in general, $Z$ no longer admits a global section, and
so is torus-fibered but not elliptically fibered. To construct
holomorphic vector bundles on the torus-fibered Calabi--Yau threefold
$Z$, one finds those bundles on $X$ which are invariant under
$\G$. These then descend to bundles on $Z$.  

We next use these results to explicitly construct a number
of three-family vacua with unification group $SU(5)$ which is
spontaneously broken to the standard gauge group 
$$
{SU(3)_{\mathrm{C}}
{\times} 
{SU(2)_{\mathrm{L}}}
{\times} 
{U(1)_{\mathrm{Y}}}}
$$
by Wilson lines
\cite{bos,W85}. This is done by taking $\G=\ZZ_2$. The restriction to
this specific class of torus-fibered threefolds and the GUT group
$SU(5)$ is for simplicity only. By the same techniques one can
construct a much wider class of torus-fibrations with more general
unification groups. This will be presented elsewhere. We also
leave a detailed exploration of the phenomenological and
cosmological aspects of these models to later publications. 

In this paper, we explicitly do the following. Sections~\ref{sec:CY}
and~\ref{sec:V} give new mathematical results necessary for
constructing semi-stable holomorphic bundles on a torus-fibered
three-fold $Z$ which admits Wilson lines. We will not give all the
derivations, but try and highlight the ideas and main results. More
mathematical details will be given in~\cite{mathpaper}. In section 2,
we discuss the general construction of $Z$, with $\p_1(Z)=\ZZ_2$, as
the quotient of elliptically fibered threefolds $X$ by an involution
$\t_X$. This leads one to consider threefolds $X$ which admit two
sections. We present the general conditions for a freely acting
involution $\t_X$ on such an $X$, and give the general homology
classes, effective curves and second Chern class of $X$. In section 3,
we construct semi-stable holomorphic vector bundles $V$ over $X$ from
a spectral cover, describing how they arise essentially by the
action of T-duality. This involves a number of new features, since the
spectral cover in such manifolds has a subtle structure. We discuss the
conditions for invariance of $V$ under the involution. This gives us a
description of bundles over $Z$. The second and third Chern classes of
these bundles are presented. 

In the last sections we return to physics. In section~\ref{sec:rules},
we give explicit rules for the construction of three-family particle
physics vacua on $Z$ with GUT group $SU(5)$. Since $\p_1(Z)=\ZZ_2$,
these vacua have Wilson lines that break $SU(5)$ to the standard $\sm$
gauge group. In section~\ref{sec:examples} we present explicit
examples of these ``standard model'' vacua for the base surfaces
$B=F_{2}$ and $dP_{3}$ of the torus fibration. Finally, we summarize
the resulting brane-world picture that emerges from these examples.

\begin{figure}[t]
   \begin{center}
      \epsfig{file=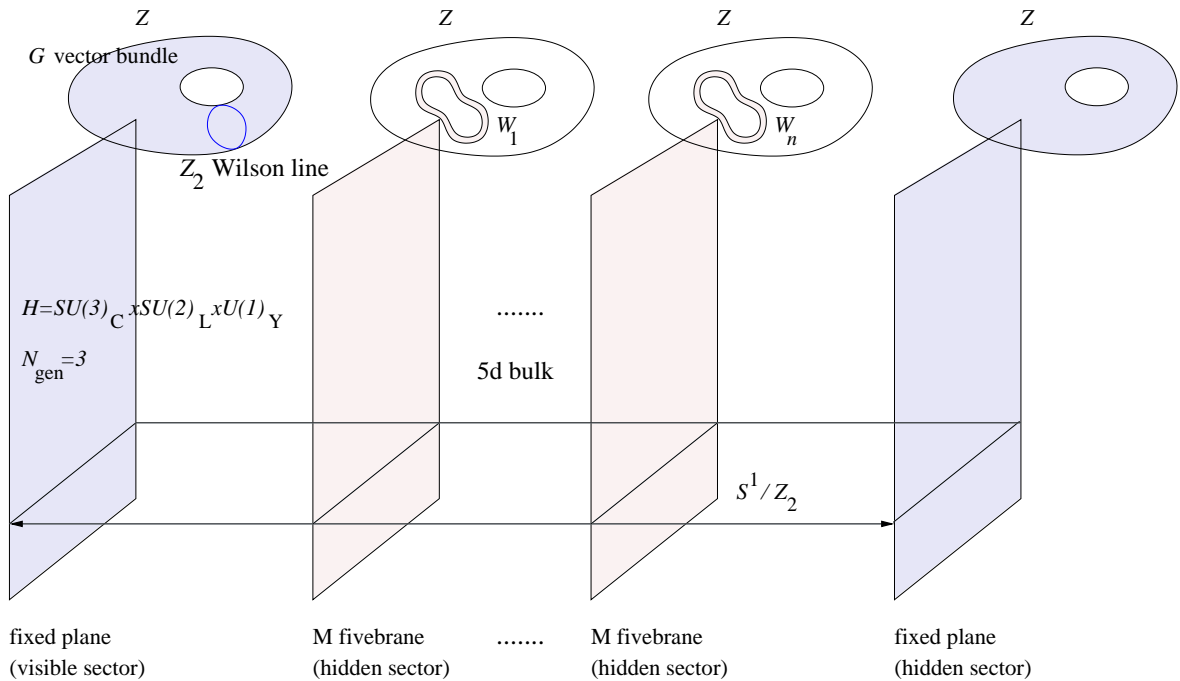,height=3.5in}
      \caption{Heterotic M-theory brane-world}
      \label{fig:braneworld}
   \end{center}
\end{figure}
In conclusion, heterotic M-theory vacua give a concrete realization of
the following ``brane-world'' (see Figure~\ref{fig:braneworld}). Six
of the eleven M-theory dimensions are compactified on a small
Calabi--Yau manifold $Z$, typically on the scale of inverse 
$10^{16}$~GeV. The additional compact direction, the $S^1/Z_2$
orbifold, is larger, perhaps of order the intermediate scale. Matter
is localized on the fixed planes of the $S^1/Z_2$ orbifold as well as
on a number of fivebranes. (The exact matter content on the fivebranes
depends on how they wrap the internal Calabi--Yau space.) In the
effective five-dimensional theory, where the spacetime is the external
$(3+1)$-dimensional space together with the $S^1/Z_2$ orbifold, the
orbifold fixed planes and the fivebranes appear as a series of domain
walls or BPS threebranes~\cite{losw1,losw2,nse}. The fixed planes
act as the boundaries of the orbifold interval, while the wrapped
fivebranes appear as a series of threebranes arrayed throughout 
$S^1/Z_2$. By choosing an appropriate gauge vacuum, one can break one
of the $E_8$ gauge groups on the orbifold fixed planes to the standard
model group $\sm$, with three families of chiral matter. Thus the
standard model is realized on one fixed plane, while the other fixed
planes and the fivebranes represent hidden sectors which only couple
gravitationally to the standard model. Finally, we should note that
related constructions, also using quotients of an elliptic Calabi--Yau
threefold, were considered in~\cite{ACK}. However, we find that on a
number of points our results are in disagreement with the calculations
and conclusions of that paper.


\section{Fibered Calabi--Yau threefolds with non-trivial $\p_1$}
\label{sec:CY}

In this section, we construct and describe the particular Calabi--Yau
manifolds $Z$ we will use to compactify Ho\v rava--Witten theory. Two
properties are required. First, we must choose particular non-trivial
gauge field configurations on the Calabi--Yau compactification to
break the $E_8$ gauge group to a smaller GUT gauge group with three
families. In this paper, we use the spectral-cover construction to
describe these gauge field configurations. This construction requires
the Calabi--Yau manifold to be elliptically fibered. The second
condition comes from breaking the GUT group down to $\sm$. This is
most naturally accomplished by including non-trivial Wilson
lines~\cite{bos,W85} on the Calabi--Yau manifold $Z$. This requires
that the threefold not be simply connected, that is, $\p_1(Z)$ must be
non-trivial.

It turns out that these two conditions are incompatible, except
possibly in one case which is discussed in \cite{don2} and also
below. Thus we are required to weaken one of the conditions and allow
$Z$ to be fibered by tori which are not necessarily elliptic. We will
review this distinction below.

To be specific, we will consider the case where the grand unified group
is 
\begin{equation}
   H = SU(5) .
\end{equation}
It is well known~\cite{bos,W85} that this can be broken to $\sm$ by a
$\ZZ_2$ Wilson line. Similar constructions work for other GUT groups, 
such as $SO(10)$ or $E_6$. Although our construction can be generalized
to larger groups, in this paper we restrict our discussion to GUT group 
$SU(5)$ and to the simplest case of finding a torus-fibered
Calabi--Yau threefold $Z$ with $\p_1(Z)=\ZZ_2$.

\subsection{Torus and elliptic fibrations}

By a \textit{torus-fibered} threefold we mean a three-dimensional
complex manifold $X$ which is equipped with a holomorphic map
\begin{equation}
   \pi: X \rightarrow B
\label{eq:1} 
\end{equation}
to some complex surface $B$ so that, for a general point $b \in B$,
the fiber $\pi^{-1}(b)$ is a smooth complex curve of genus one. The
surface $B$ is called the base of $X$. If one further demands that the
canonical bundle be trivial, then $X$ is called a torus-fibered
Calabi--Yau threefold. This condition is equivalent to assuming that 
\begin{equation}
   c_1(TX)=0 ,
\label{eq:2}
\end{equation}
where $c_{1}(TX)$ is the first Chern class of the tangent bundle $TX$.
In this paper, $X$ will always be assumed to be a smooth manifold. Note,
however, that the smoothness of $X$ does not imply the smoothness of the
base $B$. In fact, bases with isolated singularities do occur in most of 
the interesting examples below.

Now consider a holomorphic map
\begin{equation}
   \s : B \rightarrow X , 
\label{eq:3}
\end{equation}
satisfying $\pi\circ\s= \id_{B}$. Such a map is called a section of
$X$. It is important to note that a mapping of this type need not
exist. Thus, there can be torus-fibered threefolds, including
torus-fibered Calabi--Yau threefolds, that have no section. In fact,
it is precisely such Calabi--Yau threefolds that will be required to
construct realistic particle physics vacua.

Let us consider a torus-fibered threefold that does admit a
section. Then the image, $\s(B)$, of the base in $X$ intersects
each fiber $\pi^{-1}(b)$ at a unique point
$\sigma({b})$. This point plays the role of 
a natural ``zero'' for the addition
law on each fiber, turning each torus into an elliptic curve. For
this reason, a torus fibration that admits a section is called an
\textit{elliptic fibration}. An elliptic fibration naturally possesses
a line bundle, $\cL$, on $B$ whose fiber at any point $b \in B$ is the
cotangent line $T_p(\pi^{-1}(b))$ to the elliptic curve at the zero
point $p_{b}$. That is, $\cL$ is the conormal bundle to the
section $\sigma(B)$ in $X$. In the case where the elliptically fibered
manifold is a Calabi--Yau threefold, the requirement that
$c_{1}(TX)=0$ restricts the conormal bundle $\cL$ to satisfy 
\begin{equation}
   \cL = K_B^{-1} ,
\label{eq:7}
\end{equation}
where $K_B$ is the canonical bundle of the base $B$. This requirement
restricts the possible bases $B$~\cite{MV}. It turns out that, for
smooth $B$, the only possible cases are (i) the del Pezzo 
surfaces $dP_i$ for $i=1,\dots,9$, (ii) the Hirzebruch surfaces $F_r$
for any non-negative integer $r$, (iii) certain blow-ups of the
Hirzebruch surfaces and (iv) the Enriques surface $\mathcal{E}$. Thus,
elliptically fibered Calabi--Yau threefolds allow only a fairly 
restricted set of smooth base surfaces $B$. The properties of these
bases are well known and reviewed in the Appendix of~\cite{don2}. 

Elliptically fibered manifolds have a useful description in terms of a
\textit{Weierstrass model}. One recalls that a general elliptic curve
can be embedded via a cubic equation into $\PP^2$. Without loss of
generality, the equation can be put in the Weierstrass form
\begin{equation}
   zy^2 = 4x^3 - g_2 xz^2 - g_3 z^3 .
\label{eq:4}
\end{equation}
where $g_2$ and $g_3$ are general coefficients and $(x,y,z)$ are
homogeneous coordinates on $\PP^2$. To define an elliptic fibration
over a base $B$, one needs to specify how the coefficients $g_2$ and 
$g_3$ vary as one moves around the base. In general, the coefficients 
must be
sections of the line bundles $\cL^4=K_B^{-4}$ and $\cL^6=K_B^{-6}$
respectively. One notes that the torus fibers are actually elliptic 
curves, because there is always one solution to the
Weierstrass equation, namely $(x,y,z)=(0,1,0)$. This defines a global
section $\s$ of the fibration, usually called the ``zero section''. 

The elliptic curve becomes singular when two roots of the Weierstrass
equation~\eqref{eq:4} coincide. This occurs when the discriminant, 
defined by
\begin{equation}
   \Delta = g_2^3 - 27g_3^2 
\label{eq:5}
\end{equation}
vanishes. In the fibration, the discriminant is a section of
$\cL^{12}=K_B^{-12}$. The set of points in the base over which the
fibration becomes singular is given by the discriminant locus 
\begin{equation}
   \Delta = 0 .
\label{eq:6}
\end{equation}
and defines a complex curve in the base.

In some cases, although some of the elliptic fibers are singular, the
full space described by the Weierstrass model is smooth. However, as
we will see below, the Weierstrass model may also be singular. 
In this case, the
smooth Calabi--Yau threefold $X$ is a blow up of the corresponding
singular Weierstrass model.

\subsection{Involutions and the construction of $Z$ with
$\p_1(Z)=\ZZ_2$}

Unfortunately, one can show that most elliptically fibered
Calabi--Yau threefolds $Z$ have trivial $\p_1(Z)$. The one exception
is when the base $B$ is an Enriques surface. It was
shown in \cite{don2} that spectral cover constructions on 
such geometries cannot give models with three
families of chiral matter. Different constructions of bundles on such
geometries  do exist \cite{thomas} 
but they do not seem to allow enough flexibility for satisfying the
anomaly cancellation condition. Hence, we choose a different route; 
we keep the bases general
but we relax the elliptic fibration condition. 

The easiest way to produce a fibered space $Z$ with non-trivial
$\p_1(Z)$ is by quotienting an elliptically fibered space $X$ by a
freely-acting discrete symmetry. Since here we consider only the
simplest case of $\p_1(Z)=\ZZ_2$, we need only find an elliptically
fibered manifold $X$ with a freely-acting involution. Specifically,
we would like to construct an involution
\begin{equation}
   \t_X : X \to X
\end{equation}
that preserves the fibration $\pi$, as well as the holomorphic volume
form, and acts freely on $X$, that is, without fixed points. We can then
construct the quotient space
\begin{equation}
   Z = X/\tau_{X} .
\label{eq:27}
\end{equation}
Since $\tau_{X}$ acts freely and preserves the holomorphic volume
form, $Z$ is a smooth Calabi--Yau threefold. However, in general, the
section $\s$ of $X$ will not be invariant under $\t_X$, so it will not
descend to a section of $Z$. It follows that there is no reason to
expect  $Z$ to have a section and, hence, to be an elliptic
fibration. It is, in general, only a torus-fibration. 

How does one construct such an involution and what constraints does
its existence place on $X$? This is most easily analyzed by
constructing $\t_X$ as the combination of two involutions,
one acting on the base and one acting on the fiber. Note,
first, that if there is an involution $\t_X$ on $X$ that preserves the
fibration, this must project to some involution  
\begin{equation}
   \t_B : B \to B
\label{eq:21}
\end{equation}
on $B$. Since it is $X/\t_X$ that must be smooth and not
$B/\t_B$, we do not require that $\t_B$ act freely on $B$. In fact,
generally $B$ will not admit a freely-acting $\t_B$. We denote by
$\fixedset$ the set of points in $B$ that are fixed under
$\tau_{B}$. In general, $\fixedset$ can contain either a
continuous or a finite set of elements.

Fix one such involution $\t_B$. Among all Calabi-Yau manifolds $X$,
fibered over $B$, we want to show there are some for which the
involution $\t_B$ lifts to an involution $\alpha: X \to X$ which
preserves the section $\s$. We can
then combine $\alpha$ with an involution $t_\zeta$ 
on the fibers, to give $\t_X$. The
lifting is possible only if the fibration is invariant under
$\t_B$. This means that the coefficients $g_2$ and $g_3$ in
the Weierstrass equation must be invariant under $\t_B$, 
\begin{equation}
   \t_B^*(g_2) = g_2 , \qquad
   \t_B^*(g_3) = g_3 , 
\end{equation}
If these conditions are satisfied, then $\t_B$ can indeed be lifted
to an involution $\a:X \to X$, which is uniquely determined by
the additional requirements that it fix the zero section $\s$ and
that it preserve the holomorphic volume form on $X$. A local computation
shows that this $\a$ leaves fixed 
the whole fiber above each fixed point in the base, that is all
the points in $\fixedset$. Thus it is not by itself a suitable candidate
for $\t_X$.

To construct an involution without fixed points, we combine $\a$ with a
translation of the fiber. The action of translation on a smooth torus
acts without fixed points. One might then expect that a simple
translation on all the elliptic fibers of $X$ might by itself give an
involution without fixed points. The problem is that the translation
might not act freely on the singular fibers. Thus, in general, a
translation alone is not a candidate for $\t_X$ either. However, one recalls
that $\a$ left invariant only those fibers above points in the base in
$\fixedset$. If we combine $\a$ with a translation in the fiber, then,
provided none of the fibers above $\fixedset$ are singular, the
resulting transformation should act without fixed points. 

Let us be more specific. The easiest (but not most general) way to
construct an elliptically fibered Calabi--Yau threefold $X$ with a
fiber translation symmetry is to require that $X$ has two sections. 
The zero section $\sigma$ marks the zero points $p_b$ on the fibers
$\p^{-1}(b)$. Suppose there is a second section $\z$. Consider any
fiber $\pi^{-1}(b)$. Denote by $\z_{b}$ the unique point of
intersection of this fiber with the image $\z(B)$ of the base in
$X$. Let us further assume that, for any point $b \in B$,
$\z_{b}$ is  a point of order two in the fiber, that is, 
$\z_{b}+\z_{b}=\s({b})$. In terms of the sections themselves, this can
be written as  
\begin{equation}
   \z + \z = \sigma .
\label{eq:8}
\end{equation}
Using this property, one can define an involution $t_\z:X 
\rightarrow X$ as follows. Let $x$ be any point in $X$. Then $x$ lies
in a fiber $\pi^{-1}(b)$ for some $b\in B$. Define 
\begin{equation}
   t_{\z}(x) =  x + \z_{b} .
\label{eq:9}
\end{equation}
Clearly, $t_\z$ satisfies $t_\z\circ t_\z=\id_X$ and, hence, is an
involution. Furthermore, $t_\z$ has the property that 
\begin{equation}
   t_\z(\sigma) = \z , \qquad 
   t_\z(\z) = \sigma .
\label{eq:10}
\end{equation}
exchanging the two sections.

We then construct the mapping
\begin{equation}
   \t_X : X \to X ,
\label{eq:23}
\end{equation}
defined by combining $\a$ with the fiber translation $t_\z$ as 
\begin{equation}
   \tau_{X} = \alpha\circ t_{\z} .
\label{eq:24}
\end{equation}
Clearly $\t_{X}$ is an involution on $X$. By construction, $\t_{X}$
preserves the fibration $\pi$ and induces the involution $\t_{B}$ on
the base. Since $\a$ preserves the fiber, it leaves the two sections
invariant
\begin{equation}
   \alpha(\sigma) = \sigma , \qquad 
   \alpha(\z) = \z .
\label{eq:22}
\end{equation}
Combined with the action of $t_\z$ given in~\eqref{eq:10}, this
implies that $\t_{X}$ interchanges the two sections, so 
\begin{equation}
   \tau_{X}^*(\sigma) = \z , \qquad 
   \tau_{X}^*(\z) = \sigma .
\label{eq:25}
\end{equation}
Thus neither section is preserved under $\t_X$ and, consequently, the
quotient space $Z=X/\t_X$ generically has no sections and so is only
torus-fibered. Furthermore, since one can show that $t_\z$ preserves
the holomorphic volume form, we can conclude that $\tau_{X}$ also
preserves the holomorphic volume form. 

Does $\t_{X}$ have fixed points in $X$? In general, the answer is
affirmative. However, as discussed above, if none of the fibers of the
set of fixed points in the base $\fixedset$ are singular then the
action of $\t_X$ is free. Thus $\t_{X}$ will act freely on $X$ if and
only if  
\begin{equation}
   \fixedset \cap \{\Delta=0\} = \emptyset ,
\label{eq:26}
\end{equation}
where $\{\Delta=0\}$ is the discriminant locus in the base $B$.
Except for the case of the Enriques base, this then implies that 
$\fixedset$ must consist of a finite number of fixed points.

Furthermore, since $\tau_{X}$ preserves the fibration $\pi$ of $X$ and
induces the involution $\tau_{B}$ on $B$, it follows that $Z$ is a
torus-fibration over the base space  
\begin{equation}
   S = B/\tau_{B} .
\label{eq:28}
\end{equation}
Recall that, since involution $\tau_{B}$ generically has
fixed points, $S$, unlike $Z$, is generically not a smooth manifold. 

In summary, we can construct a torus-fibered Calabi--Yau threefold
with $\p^1(Z)=\ZZ_2$ by the following quotient
\begin{equation}
\begin{CD}
       X     @>q>>   Z=X/\t_X \\
   @V{\pi}VV          @VVV    \\
       B     @>>>    S=B/\t_B
\end{CD} 
\label{eq:29}
\end{equation}
In order to construct such a manifold $Z$ we require
\begin{itemize}
\item $X$ has two sections $\s$ and $\z$ such that $\s+\s=\z$ under
   fiber-wise addition,
\item $\t_B$ acts on the base $B$ with a fixed point set $\fixedset$
   with the property that $\fixedset\cap\{\D=0\}=\emptyset$,
\item in the Weierstrass model for $X$ we have $\t_B^*(g_2)=g_2$ and
   $\t_B^*(g_3)=g_3$, ensuring that the involution preserves the fibration.
\end{itemize}
In the next subsection, we will discuss what these conditions imply
about the structure of $X$. 

Finally, we note that the Chern classes 
$c_i \in H^*(Z,{\mathbb{Q}}) $
of the tangent bundle $TZ$ can be determined 
from the Chern classes of $TX$ as follows. Let 
\begin{equation}
   q:X \rightarrow Z
\label{eq:29A}
\end{equation}
be the quotient map. Since $X$ is a double cover of $Z$, it follows
that 
\begin{equation}
   c_{i}(TZ) = \frac{1}{2}q_{*}c_{i}(TX) ,
\label{eq:29B}
\end{equation}
where $q_{*}c_i(TX)$ is the push-forward of $c_i(TX)$.

\subsection{Structure of $X$} 

The main requirement on $X$ is that it is a smooth elliptically
fibered Calabi--Yau threefold admitting two sections. Such manifolds
can be constructed from the corresponding Weierstrass model as
follows. First, in order to have a pair of sections $\sigma$ and
$\z$, the Weierstrass polynomial must factorize as
\begin{equation}
   zy^2 = 4(x-az)(x^2 + axz + bz^2) ,
\label{eq:11}
\end{equation}
where, comparing to equation \eqref{eq:4}, we see that
\begin{equation}
   g_2 = 4(a^2-b) , \qquad 
   g_3 = 4ab .
\label{eq:12}
\end{equation}
Note that this implies that $a$ and $b$ are sections of $K_{B}^{-2}$ and 
$K_{B}^{-4}$ respectively.
The zero section $\sigma$ is given by $(x,y,z)=(0,1,0)$ and the second
section $\z$ by $(x,y,z)=(a,0,1)$. The fibers are singular over the
discriminant curve $\Delta=0$, where
\begin{equation}
   \Delta = \Delta_{1}\Delta_{2}^{2}
\label{eq:13}
\end{equation}
and
\begin{equation}
   \Delta_{1} = a^{2}-4b , \qquad  
   \Delta_{2} = 4(2a^{2}+b) .
\label{eq:14}
\end{equation}
We see that the discriminant curve has two components. 

\begin{figure}[ht]
   \begin{center}
      \epsfig{file=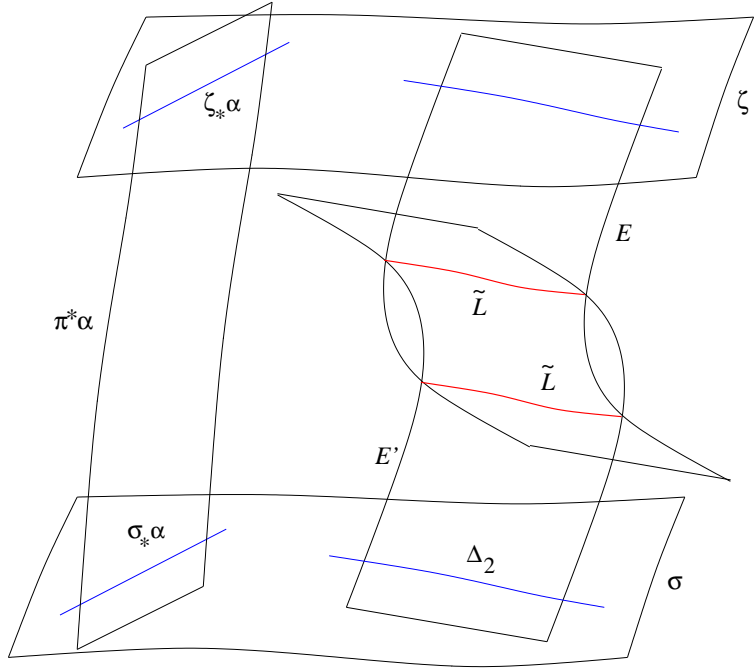,height=3.5in}
      \caption{Structure of $X$}
      \label{fig:cy}
   \end{center}
\end{figure}

As constructed, the Weierstrass model is singular. It is easy to show
that there is a curve of singularities over the $\D_2=0$ component of
the discriminant curve. We note that the vanishing of $\Delta_{2}$
corresponds to one of the roots of the second factor of the
Weierstrass polynomial \eqref{eq:11} being coincident with the zero of
the $x-az$ factor. Consequently, the singular points over the part of
the discriminant curve where $\Delta_{2}=0$ all live in the $\z$
section. Specifically, the singular points form a curve $L$ in section
$\z$ given by  
\begin{equation}
   (x,y,z) = (a,0,1) , \qquad 
   2a^{2} + b = 0 ,
\label{eq:15}
\end{equation}
To construct the smooth Calabi--Yau threefold, it is necessary to
blow up this entire curve. (It turns out that one blowup suffices.)
This is achieved as follows. The singular
point of each individual fiber over $\Delta_{2}=0$ is replaced by a
sphere $\PP^{1}$. This is a new curve in the Calabi--Yau threefold,
which we denote by $N$. This reflects the fact 
that the general elliptic fiber
$F$ has split over $\D_2=0$ into two spheres: the new fiber $N$
plus the proper transform of the singular fiber, which is in the class
$F-N$. The union of these new fibers over the curve of
singularities~\eqref{eq:15} forms a surface in the Calabi--Yau
threefold, specifically, an exceptional divisor denoted by $E$. There
is an analogous surface $E'$ formed by the union of the $F-N$ fibers
over $\D_2$ in the zero section. The blown-up Weierstrass model is the
smooth elliptically fibered Calabi--Yau threefold $X$ with two
sections. The blown up space together with the inverse image $\tilde{L}$ of
the singular curve $L$ is shown in Figure~\ref{fig:cy}.

We recall that to admit an involution we required in addition that
$g_2$ and $g_3$ were invariant under $\t_B$. In terms of the
parameters $a$ and $b$ this translates into 
\begin{equation}
   \t_B^*(a) = a , \qquad
   \t_B^*(b) = b .
\label{eq:abinv}
\end{equation}
The action of the involution $\t_X$ can be seen in
Figure~\ref{fig:cy}. We recall that, under $\t_X$, the two sections
are exchanged. Similarly, the two surfaces $E$ and $E'$ are
exchanged. A generic fiber $F_b$ over a point $b\in B$, will be
exchanged with the fiber $F_{b'}$ over the image point $b'=\t_B(b)$
under the involution on $B$.

\subsection{Classes, effective curves and the Chern class $c_{2}(TX)$}
\label{Xclasses}

Let us consider the smooth elliptically fibered Calabi--Yau threefold
$X$ with two sections, constructed as in the previous section. We
would like to identify the independent classes of both surfaces and
curves on $X$. For the classes of surfaces, generically, we have the
classes of the two sections $\sigma$ and $\z$, of the new exceptional
divisor $E$ obtained from the blowing-up procedure and the pull-back
to $X$ of classes of curves $\alpha$ from the base $B$, which we
denote by $\pi^{*}\alpha$. In fact, these divisor classes are not all
independent. For the classes of curves, generically, we have the class
of elliptic fiber $F$, of the new class $N$ obtained from the
blowing-up procedure, the image $\tilde{L}$ of the curve of
singularities and the embeddings via the two sections of classes of
curves $\alpha$ from the base $B$, which we write as $\sigma_*\alpha$
and $\z_*\alpha$ respectively. Once again, these curve classes
are not all independent. An independent set of divisors and curves, 
together with their intersections, is given by 
\begin{equation}
\begin{array}{c|ccc}
   & \p^*\a & \s & \z \\
   \hline
   \s_*\b & \a\cdot\b & -c_1\cdot\b & 0 \\
   F-N & 0 & 1 & 0 \\
   N & 0 & 0 & 1 \\
\end{array} ,
\label{eq:16}
\end{equation}
where $\alpha$ and $\beta$ are classes of curves in $B$ and $c_{1}$
denotes the first Chern class of the base, $c_{1}(B)$. 

The exceptional divisor from the blowing up procedure is given by
\begin{equation}
   E = 2 \left(\s - \z + \p^*c_1\right).
\label{E}
\end{equation}
while it image $E'$ under $\t_X$ is 
\begin{equation}
   E' = 2 \left(\z - \s + \p^*c_1\right).
\end{equation}
Meanwhile, the transform of the curve of singularities is 
\begin{equation}
   \tilde{L} = 8 [\s_*c_1 + c_1^2 (F-N)]
\label{Ltilde}
\end{equation}
and we have the important relationship between classes of curves
\begin{equation}
   \z\cdot\p^*\a = \s\cdot\p^*\a + (\a\cdot c_1)[(F - N) - N].
\label{zpa}
\end{equation}
Finally, to complete the description of the cohomology ring we have the
following intersections between divisors
\begin{equation}
\begin{gathered}
   \s\cdot\s = -\s_*c_1 , \qquad
      \z\cdot\z = -\z_*c_1 , \\
   \s\cdot\z = 0 , \\
   \s\cdot\p^*\a = \s_*\a , \qquad
      \z\cdot\p^*\a = \z_*\a
\end{gathered}
\end{equation}

The action of the involution $\t_X$ on each class is easy to
identify. We have for the divisors
\begin{equation}
\begin{gathered}
   \t_X^*(\s) = \z , \qquad  \t_X^*(\z) = \s , \\
   \tau_X^*(\p^*\a) = \p^*\t_B(\a) ,
\end{gathered}
\label{tXdivisors}
\end{equation}
while for the curves
\begin{equation}
\begin{gathered}
   \t_X^*(F-N) = N , \qquad \t_X^*(N) = F-N , \\
   \t_X^*\s_*\a = \z_*\t_B(\a) = \s_*\t_B(\a) + (\a\cdot c_1)[(F-N) - N], \\
   \t_X^*\z_*\a = \s_*\t_B(\a)
\end{gathered}
\label{tXcurves}
\end{equation}

When describing the fivebranes in our model, it is important to
identify the \textit{effective} classes of curves in $X$. As
discussed in~\cite{don1,don2,don3}, these are the cohomology classes
which can be realized by the sum curves in the Calabi--Yau
corresponding to an actual collection of fivebranes (and no
anti-fivebranes). Not all classes are of this form. Thus we would like
to identify the conditions for a class $[W]$ in $X$ to be effective,
that is, for $[W]$ to lie in the Mori cone of $H_{2}(X,\ZZ)$. We
recall that in blowing up the Weierstrass model, the singular point of
a fiber over the $\Delta_{2}=0$ was replaced by a sphere $\PP^{1}$
represented by the new class $N$. This class is by definition
effective. Recall further that the fiber over $\D_2=0$ in fact splits
into a pair of spheres. The other sphere is the proper transform of
the singular fiber and is in the class $F-N$. It is by definition
effective. Furthermore, the lift $\s_*\a$ of any effective class $\a$
in the base into the zero section $\s$ is must also be effective in
$X$. It follows from \eqref{eq:16} that we can, in general, write the
class of a curve $W$ in $H_{2}(X,\ZZ)$ as   
\begin{equation}
   [W] = \s_*\o + c(F-N) + dN ,
\label{eq:18}
\end{equation}
where $\o$ is a class in the base $B$ and $c$ and $d$ are integers. 
It is easy to show that 
a sufficient condition for $W$ to be an effective class is that
\begin{equation}
   \omega \text{ is effective in $B$}, 
       \quad c \geq 0, \quad d \geq 0 .
   \quad \implies \quad
       [W] \text{ is effective in X}  
\label{eq:19}
\end{equation}
We will, in the following, often denote an expression such as ``$[W]$
is effective in $X$'' simply by $[W]\geq0$. 

It will be essential in this paper to know the second Chern class of
the tangent bundle $TX$ of the relevant Calabi--Yau threefolds. The
second Chern class $c_{2}(TX)$ of an elliptically fibered Calabi--Yau
threefold with two sections can be calculated by blowing up the
associated Weierstrass model, as discussed above. Here, we simply
present the result. We find that 
\begin{equation}
   c_2(TX) = 
      12\s\cdot\pi^*c_1 + (c_2+11c_1^2) (F-N) + (c_2-c_1^2) N ,
\label{eq:20}
\end{equation}
where $c_1$ and $c_2$ stand for the Chern classes $c_1(B)$ and $c_2(B)$ 
of the base $B$ respectively. This formula was originally given,
though in a different form, by Andreas, Curio and
Klemm~\cite{ACK}. Note that setting $N$ to zero reduces expression
\eqref{eq:20} to the original formula given by Friedman, Morgan and
Witten~\cite{FMW} for the case of elliptically fibered Calabi--Yau
threefolds with a single section $\sigma$.


\section{Holomorphic vector bundles on fibered Calabi--Yau threefolds}
\label{sec:V}

Having identified the structure of the torus-fibered Calabi--Yau
manifold $Z$ on which we will compactify, the next question is to
construct general supersymmetric gauge vacua on $Z$, in particular, 
vacua with structure group $G=SU(5)\subset E_8$. As discussed in
the introduction, constructing supersymmetric vacua corresponds to
constructing general semi-stable holomorphic vector bundles over
$Z$. We recall that the work of~\cite{FMW,D,BJPS} shows how to
construct such bundles over elliptically fibered manifolds. Here 
$Z$ is the quotient $X/\t_X$ of an elliptically fibered threefold $X$
by an involution $\t_X$. Thus the easiest way to construct bundles
over $Z$ is to construct suitable vector bundles $V$ over
$X$ which descend to the quotient threefold $Z$. This restricts us to
those bundles on $X$ which are invariant under the involution. That
is, we will need to find $V$ such that $\tau_X^*(V)=V$. 

In this section, we discuss the details of such a construction. We
first review the spectral cover construction of semi-stable
holomorphic bundles $V$ over an elliptically fibered manifold $X$. We
stress a physical picture, where the spectral data can be viewed as
the T-dual of the gauge bundle $V$ over $X$. We then discuss the
structure of the spectral data, in the case of the particular class of
Calabi--Yau manifolds with two sections discussed in the previous
section. Finally, we turn to the question of identifying those $V$
which are invariant under the involution. We also give explicit
expressions for the Chern classes of $V$ and for the corresponding
bundles on $Z$. A number of important subtleties complicate this
analysis as compared with the generic case where $X$ has only one
section. For simplicity, we will not give the derivations of all of our
results in this paper. Also, we will often use the Chern classes to
characterize a bundle rather than deal with the bundles
themselves. The full mathematical details will be given
elsewhere~\cite{mathpaper}.

\subsection{T-duality and the spectral construction}

We would like to characterize semi-stable holomorphic $SU(n)$ vector
bundles over an elliptically fibered Calabi--Yau threefold
$X$. Let us generalize slightly and consider $U(n)$ bundles
of degree zero over the fibers. It was shown in~\cite{FMW,D,BJPS}
that, generically, a semi-stable rank $n$ holomorphic vector bundle
$V$ on $X$ can be constructed from two objects  
\begin{itemize}
\item 
   a divisor $C$ of $X$ which is an $n$-fold cover of the base
   $B$, known as the spectral cover, 
\item 
   a line bundle $\cN$ on $C$.
\end{itemize}
The relationship between the spectral data $(C,\cN)$ and the bundle
$V$ actually has a simple physical interpretation. The spectral data
is the ``T-dual'' of the semi-stable holomorphic bundle $V$. 

To understand the what is meant by T-duality in this context, recall
that the Calabi--Yau manifold $X$ has an elliptically fibered
structure. This means that we can consider making a T-duality
transformation on the fiber. Specifically, on each fiber we make a
T-duality transformation along each cycle of the torus, preserving the
complex structure on the torus, but inverting the volume. Of course,
the transformation is more subtle at the points where the fibration is
singular, but it can in general be defined. The resulting T-dual
manifold $\tX$, is also elliptically fibered. In fact, with regard to
complex structure, it is isomorphic to the original manifold
$X$. Mathematically, it is the ``associated Jacobian bundle''
$\tX\equiv J(\p)$. Thus we have
\begin{equation}
\begin{CD}
   X @>{\text{T-duality}}>> \tX \equiv J(\p) \cong X
\end{CD}
\end{equation}
by the action of T-duality on each fiber. 

What happens to the vector bundle $V$ under such a duality
transformation? Physically, we recall that vector bundles (or, more
generally, sheaves) correspond to a collection of
D-branes. The low-energy theory of $n$ coincident D-branes
is a $(D+1)$-dimensional Yang--Mills theory with a gauge group
$U(n)$. From this point of view, we can view our rank $n$ vector
bundle on the threefold $X$ as $n$ D6-branes all wrapping the
Calabi--Yau manifold. If $V$ is non-trivial, the configuration is 
a source not only of D6-brane charge but also of D4-brane, D2-brane
and D0-brane charge. The charges are given~\cite{Dcharge} in terms of the
Chern character classes $\ch_0(V)=n$, $\ch_1(V)$, $\ch_2(V)$ and
$\ch_3(V)$ for D6-, D4-, D2- and D0-brane charge respectively. Thus,
in general, the bundle describes some collection D6-, D4-, D2- and
D0-branes. 

This description can be extended to the case where we have, for
instance, only D4-, D2- and D0-branes on $X$. This implies that the rank
of the bundle is zero (no D6-brane charge) but the the higher Chern
classes are non-zero. This is, of course, not possible to describe in
terms of a vector bundle, but can be described as a ``sheaf''. A simple
example is a single D4-brane wrapping a divisor $C$ in $X$. There 
will be a line-bundle $\cN$ on $C$ describing the $U(1)$ gauge fields
on $X$ (and in general describing embedded D2- and D0-brane
charge). There is an inclusion map $i_C:C\to X$. 
We can try and use this map to push the bundle on $C$ into an object
on $X$. This will be a ``bundle'' $i_{C*}\cN$ which is simply $\cN$
when restricted to the D4-brane on $C$ but everywhere else in $X$ the
fiber is dimensionless. Such an object is a sheaf. A general vector
bundle can then be regarded as a special type of sheaf. 

Viewing $V$ as a set of D-branes, we know that D-branes are in general
mapped under T-duality to new D-branes of different
dimension. Roughly, D4-, D2- and D0-branes transverse to the fiber of
$X$ should become D6-, D4- and D2-branes wrapping the fiber of $\tX$,
while D6-, D4- and D2-branes wrapping the fiber of $X$ become D4-, D2-
and D0-branes transverse to the fiber of $\tX$. (The actual
transformation is more complicated because of the curvature of $X$ and
the singular fibers, but this is the general picture.) Thus, in
general, the bundle $V$ should map to some new bundle (or more
generally new sheaf) $\tV$ on $\tX$ describing the T-dual
configuration of D-branes. Since we can dualize back from $\tX$ to
$X$, the map is invertible. In what follows, it is more convenient to
define the map in terms of the duality from the bundle $\tV$ on $\tX$
to $V$ on $X$. Thus, T-duality induces an invertible map $S$ between
sheaves $\tV$ on $\tX$ and sheaves $V$ on $X$ so that  
\begin{equation}
   V = S (\tV) .
\end{equation}
Mathematically, this map is known as the ``Fourier--Mukai''
transform. (To be more precise, this transform acts not on sheaves 
but on the derived category of sheaves on $X$ as described, for
instance, in~\cite{AD}.)  We note that it induces a map $s$ on
cohomology exchanging the D-brane charges: 
\begin{equation}
   \ch_i (V) = s \; \ch_i (\tV)
\label{smap}
\end{equation}
We would expect that the T-duality acts linearly on the D-brane
charges. And indeed, one can show that $s$ is linear on cohomology. 

\begin{figure}[ht]
   \begin{center}
      \epsfig{file=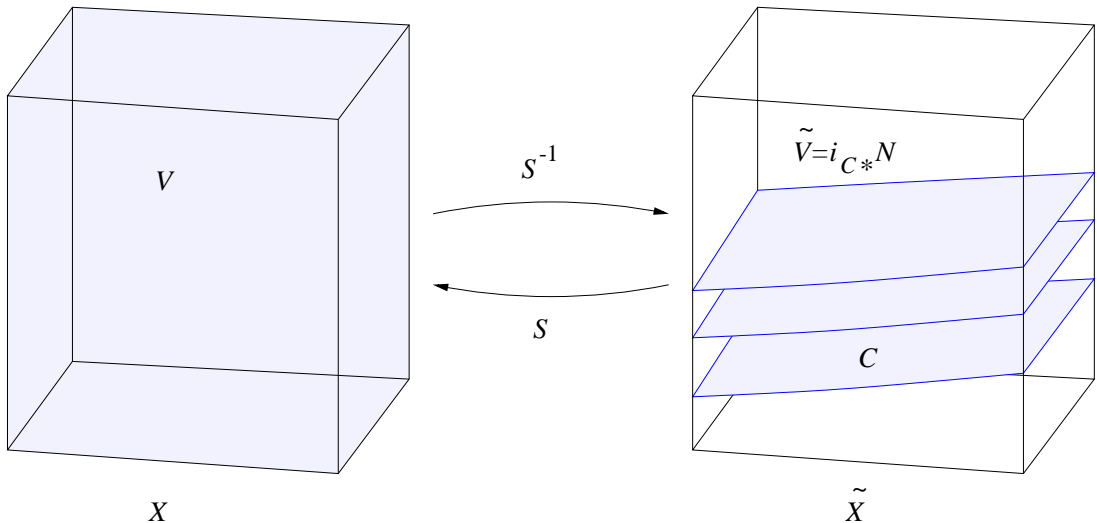,height=2.7in}
      \caption{The Fourier--Mukai T-duality transform}
      \label{fig:FMtransform}
   \end{center}
\end{figure}

So far, while interesting, the T-duality map has not
helped us solve the problem of characterizing semi-stable bundles $V$
on $X$. In general, the dual bundle $\tV$ appears to be just as
complicated as the original bundle $V$. However, we have not yet
utilized the semi-stability condition. The spectral cover
construction makes a specific semi-stability assumption, namely that
the bundle is still semi-stable when restricted to
a generic fiber
\footnote{It is known \cite{FMW} that for an
appropriate choice of the K\"ahler form $\o$ on $X$, $o$-stability
indeed implies semi-stability on the generic fiber.}. 
The consequence of this is that if $V$ is
semi-stable then its T-dual $\tV=S^{-1}(V)$ describes only
D4-branes and no D6-branes. In particular, this is precisely the
information contained in the spectral data. Namely, the spectral cover
$C$ is the surface in $\tX$ on which the D4-brane wraps and $\cN$
describes the gauge field on the brane. Thus, formally, if 
\begin{equation}
   i_C : C \to \tX 
\label{iC}
\end{equation}
is the inclusion map of the spectral cover, the T-dual $\tV$ of $V$ is
the sheaf
\begin{equation}
   \tV = i_{C*} \cN .
\end{equation}
This then is the spectral cover construction
\begin{itemize}
\item the bundle $V$ is the fiber-wise T-dual or ``Fourier--Mukai
   transform'' 
   \begin{equation}
      V = S(i_{C*}\cN)
   \label{FM}
   \end{equation}
   of the sheaf $\tV=i_{C*}\cN$ as defined by the spectral data
   $(C,\cN)$. 
\end{itemize}
We note that the spectral data naturally lives in the isomorphic
T-dual space $\tX$. The construction is shown diagrammatically in
Figure~\ref{fig:FMtransform}.  

One can get an idea of why the semi-stability condition implies that
$\tV$ contains no D6-branes as follows. It is easy to show that
requiring the restriction of $V$ to a generic fiber to be semi-stable
implies that first Chern class $c_1(V)$ restricted to a
generic fiber is zero. This clearly restricts the form of the D4-brane
charge. In particular, it means that there can be no D4-branes
wrapping sections of $X$. However, it is precisely D4-branes wrapped on
sections of $X$ which become D6-branes under T-duality since they are
transverse to the fiber direction. Thus $\tV$ contains only D4-, D2-
and D0-branes. Note that, in addition, the original $n$ D6-branes
will map into $n$ D4-branes wrapping sections of the dual $\tX$. To be
more precise, they in general map to a D4-brane which is an $n$-fold
cover of the base $B$. This is indeed precisely the form of the
spectral cover $C$.

\subsection{Bundles on $X$ and the form of the spectral data} 

Let us now return to the specific question of the description of
semi-stable holomorphic $SU(n)$ bundles on $X$. We will find that the
specific form we chose for $X$ gives additional 
degrees of freedom for the spectral data, which are
not present in the generic case. In particular, as we discuss, the
spectral cover is no longer everywhere a finite cover of the base
$B$. In this paper, we will mention some of the complexities this
introduces, but save the more mathematical analysis for a later
publication~\cite{mathpaper}. 

First, we need to impose the condition that we have an $SU(n)$
bundle, so that 
\begin{equation}
   c_{1}(V) = 0 ,
\label{eq:31}
\end{equation}
where $c_{1}(V)$ is the first Chern class of the bundle $V$. The map
$s$ in cohomology induced by the Fourier--Mukai transform~\eqref{smap}
gives expressions for the Chern classes $c_i(V)$ in terms of the Chern
classes $c_i(\tV)$ of $\tV=i_{C*}\cN$. The latter are easily expressed
in terms of $C$ and the first Chern class $c_1(\cN)$ of the line
bundle $\cN$. We will not give the details here but simply present the
results. Requiring that $c_1(V)$ vanishes on $X$, rather than just on
the fibers, imposes restrictions on $C$ and $\cN$. We find that the
class of the spectral cover is constrained to be 
\begin{equation} 
   C = n\sigma + \pi^{*}\eta ,
\label{eq:32}
\end{equation}
where $\p^*\eta$ is the pull-back into $X$ of some divisor class
$\eta$ in the base $B$. Since $C$ is an actual surface in $\tX$, we
have the addition condition that 
\begin{equation}
   \text{$\eta$ must be an effective class on $B$}
\end{equation}
Note, in addition, that equation \eqref{eq:32} involves only
the zero section $\s$ and not the second section $\z$. Furthermore,
the vanishing of $c_1(V)$ also requires that 
\begin{equation} 
   c_1(\cN) = \hf \left.(C+\p^*c_1)\right|_C + \g ,
\label{eq:33}
\end{equation}
where the first term is the restriction to $C$ of a divisor class in
$X$, while $\g$ is a class on $C$ with the property that, if $p$ is the
projection $p:C\to B$, then the push-forward of $\g$ vanishes
\begin{equation} 
   p_* \g = 0
\end{equation}
as an element of $H_2(B,\ZZ)$.

We would like to have an explicit expression for the additional
freedom $\g$ in the line bundle $\cN$. To do this, we need to know the
generic classes on the spectral cover $C$ and then find which of these
classes satisfy the projection condition $p_*\g=0$. 

If $C$ were completely generic, we would expect that the only classes on
$C$ are those that come from the restriction of classes in $X$
to $C$. That is, we expect classes of the form $\g=D|_C$ where
$D\in H_4(X,\ZZ)$ is a divisor class on $X$. From
section~\ref{Xclasses}, the divisors on $X$ are of the form
$D=a\s+b\z+\p^*\a$ where $\a$ is a class in $H_2(B,\ZZ)$. However, we
recall that the $c_1(V)=0$ condition meant that the spectral cover was
not generic, but was restricted so that the class of $C$ is of the form
$n\s+\p^*\eta$. In this case, extra classes appear on the spectral
cover. They can be seen by considering the intersection of $C$ with
the exceptional divisor $E$. We have from~\eqref{eq:16} and~\eqref{E}
that, as a class in $X$,
\begin{equation}
   E|_C = E \cdot C = 2(\s - \z + \p^*c_1)\cdot(n\s + \p^*\eta)
      = 4(\eta\cdot c_1)N
\end{equation}
Since both $C$ and $E$ are actual surfaces in $X$, this implies that
they intersect in a curve wrapping the new fiber $4\eta\cdot
c_1$ times. For generic $C$, if we project down to the base, the
curve $C\cdot E$ projects to $4\eta\cdot c_1$ distinct points. Thus
we can argue that, generically, the curve $C\cdot E$ splits into
$4\eta\cdot c_1$ distinct curves each wrapping the new 
component of the fiber over a
different point in the base. In $X$, all of these curves are in the
same homology class. However, in $C$, since each curve can be
separately blown down, they must be in distinct cohomology classes in
$H_2(C,\ZZ)$. Let us denote these distinct classes as $N_i$ for
$i=1,\dots,4\eta\cdot c_1$. Thus, as a class in $C$, we have
$E|_C=\sum_i N_i$. 

The fact that $C$ necessarily includes the exceptional curves $N_i$
has important consequences. In particular, it means that $C$ is no
longer everywhere a finite cover of $B$. Generically it is an $n$-fold
cover, but there are $4\eta\cdot c_1$ special points in $B$ where it
spans the whole of a new fiber $N$ above the base. This has important
consequences for the spectral cover construction. In the previous
section, we claimed that a smooth semi-stable holomorphic vector
bundle $V$ on $X$ is the Fourier--Mukai transform of the sheaf
$i_{C*}\cN$ on $\tX$, where $\cN$ is a line-bundle on $C$. However,
this is only strictly true for generic $C$. If $C$ is singular or is
no longer a finite cover, more general sheaves on $\tX$ can lead to
a smooth semi-stable bundle $V$. In particular, one is led to consider
sheaves supported on the new fibers $N_i$. We leave the discussion of
these subtleties to a more mathematical paper~\cite{mathpaper}. In
this paper, we will restrict ourselves to the simple case of line bundles
over $C$, and will mostly been concerned with the Chern classes rather
than the specific realization of the bundles themselves. 

Returning to the classes on $C$, in summary, for a generic spectral
cover $C=n\s+\p^*\eta$ in $X$, the generic classes on $C$ can be taken
as
\begin{equation}
\begin{aligned}
   \text{restrictions from $X$: }\qquad &
      \s|_C, \quad \p^*\a|_C , \\
   \text{new fiber classes: }\qquad &
      N_i ,
\end{aligned}
\end{equation}
where $i=1,\dots,4\eta\cdot c_1$. Furthermore, from their
intersections in $X$ and the fact that each $N_i$ can be separately
blown down, one can show that 
\begin{equation}
   N_i \cdot N_j = -2 \d_{ij}
\end{equation}
as an intersection in $C$, while
\begin{equation}
   N_i \cdot \s|_C = N_i \cdot \p^*\a|_C = 0 ,
\end{equation}
for all $i$. One might expect that there is also a class from
$\z|_C$. However, we recall that $E=2(\s-\z+\p^*c_1)$, so this class is
not independent. We note, however, that $\hf E$ is also an integral
class on $X$. Thus, we do have the useful fact that  
\begin{equation}
   \hf E|_C = \hf \sum_i N_i
\end{equation}
is actually an integral class on $C$, despite the rational
coefficients in the sum. 

Having identified the classes on $C$, we can now find the generic form
of $\g$. A short calculation shows that $\g$ has the form
\begin{equation}
   \g = \l \left(n\s - \pi^*\eta + n \pi^*c_1 \right)\cdot C 
              + \sum_i \k_i N_i .
\label{eq:34}
\end{equation}
In order for $c_{1}(\mathcal{N})$ to be an integral homology class,
the coefficients $\l$ and $\k_i$ are restricted, so that, in the case where
$n$ is odd, we have, recalling that $\hf\sum_iN_i$ is integral
\begin{equation}
\begin{aligned}
   \l - \shf &\in \ZZ  , \\
   \k_i - \shf m &\in \ZZ ,
\end{aligned}
\label{eq:35}
\end{equation}
where $m$ is an integer. In this paper, we restrict our results to $n$
odd since all of the examples we present below will involve the
structure group $SU(5)$. 

The other Chern classes of $V$, which by the Fourier--Mukai transform
could be written in terms of $C=n\s+\p^*\eta$ and $\cN$, then reduce
to expression in terms of $\eta$, $\l$ and $\k_i$. We find that 
\begin{equation}
\begin{aligned}
   c_2(V) =& \s\cdot\p^*\eta 
       - \left\{ \frac{1}{24}\left(n^3-n\right)c_1^2 
            - \hf\left(\l^2-\frac{1}{4}\right)n
                \eta\left(\eta-nc_1\right) 
            - \sum_i\k_i^2 
            \right\} (F-N) \\ &
       - \left\{ \frac{1}{24}\left(n^3-n\right)c_1^2 
            -\frac{1}{2}\left(\l^2-\frac{1}{4}\right)n
                \eta\left(\eta-nc_1\right) 
            - \sum\k_i^2 + \sum_i\k_i 
            \right\} N ,
\end{aligned}
\label{eq:36}
\end{equation}
and 
\begin{equation}
   c_3(V) = 2 \l \eta \left(\eta-nc_1\right) ,
\label{eq:37}
\end{equation}
where $c_1$ and $c_2$ stand for the Chern classes $c_1(B)$ and
$c_{2}(B)$ of the base $B$ respectively. 
It is important to note that the expression for $c_2$ depends on
the $\k_i$, but the one for $c_3$ does not. Setting $\k_i$ to
zero in \eqref{eq:36} returns one to the formula for $c_2$ given by
Friedman, Morgan, Witten~\cite{FMW}, as it must. 
The single section threefolds case of Expression
\eqref{eq:37} for $c_3(V)$ was derived in \cite{curio}.

\subsection{Vector Bundles on $Z=X/\t_X$}

We now want to construct semi-stable holomorphic vector bundles $V_Z$
on the torus-fibration $Z=X/\tau_{X}$. Every such bundle can be pulled
back to a bundle $V$ over $X$ that is invariant under the involution
$\tau_{X}$ and conversely. That is, if 
\begin{equation}
   \tau_{X}^*(V)=V 
\label{eq:38}
\end{equation}
then $V$ is the pull-back of a bundle $V_{Z}$ on $Z$. We therefore
must characterize the condition that $V$ be invariant under
$\tau_{X}$. 

Recall that $V$ is constructed as the Fourier--Mukai transform of the
sheaf $\tV=i_{C*}\cN$ given the spectral cover $C$ and the line bundle
$\cN$ on $C$. The generic forms of $C$ and $\cN$ are given in
equations \eqref{eq:32} and \eqref{eq:34} respectively. What
additional conditions on the spectral data does the
requirement~\eqref{eq:38} that $V$ is invariant under $\t_X$ imply? 

This can be calculated directly by finding the induced action of
$\t_X$ on the spectral data. In particular, if $S$ is the
Fourier--Mukai transform from $\tV$ to $V$, and $S^{-1}$ the inverse
transform, we have that under $\t_X$ 
\begin{equation}
   \tV = i_{C*}\cN \to \tV' = S^{-1} \circ \t_X^* \circ S (i_{C*}\cN)
\end{equation}
Rather than considering this action in general, let us start by
considering the condition invariance will place on the spectral cover,
and then on the Chern classes of $\tV$. 

We recall that the spectral data live naturally in the T-dual space
$\tX$. Furthermore, the involution $\t_X$ is the combination of two
involutions, $\t_X=\a\circ t_\z$, where $\a$ is the lift of the
involution $\t_B$ and $t_\z$ is a translation in the fiber of
$X$. Under T-duality, a translation on the fiber of $X$ has no
realization as an action on the dual space $\tX$ so the induced
action of $t_\z$ on $\tX$ is trivial. The involution on the base
$\t_B$, however, carries over into the corresponding involution in the
base of $\tX=J(\pi)$. Thus, recalling that $X$ and $\tX$ are
isomorphic, we expect that one of the requirements of the invariance
of $V$ under $\t_X$ is that the T-dual D4-brane described by the
spectral cover $C$ is invariant under $\a$. That is, we require that 
\begin{equation}
   \a(C) = C 
\label{eq:39}
\end{equation}
Recall from \eqref{eq:22} that $\s$ is invariant under $\a$. We find then,
using the form of $C$ given in equations \eqref{eq:32}, that the
condition \eqref{eq:39} is satisfied provided  
\begin{equation}
   \tau_{B}(\eta) = \eta .
\label{etainv}
\end{equation}

It is more subtle, however, to calculate the induced transformation of
$\cN$ under the action of $\t_X$ on $V$. Here, rather than calculate
the action of $S^{-1}\circ\t_X^*\circ S$ directly, we will consider the 
transformation of the Chern classes of $V$ to get information about
how $\cN$ transforms. Let us write $V'=\t_X^*(V)$ for the pullback of
the bundle $V$ under the involution and $\eta'$, $\l'$ and $\k_i'$ for
the corresponding spectral data. Given the transformation of the
classes on $X$ under $\t_X$ given in~\eqref{tXdivisors}
and~\eqref{tXcurves}, we find that, assuming the spectral cover
condition~\eqref{etainv} that $\eta'\equiv\t_B^*(\eta)=\eta$,  
\begin{equation}
\begin{aligned}
   c_2(V') = \t_X^*(c_2(V)) &= 
        c_2(V) + \sum_i (\shf - \k_i) \left[N - (F-N)\right] \\
   c_3(V') = \t_X^*(c_3(V)) &= c_3(V)
\end{aligned}
\end{equation}
Comparing this to the general expressions~\eqref{eq:36}
and~\eqref{eq:37} for $c_2(V)$ and $c_3(V)$, we can conclude that under
$\t_X$ we must have that $\l$ remains unchanged. 
\begin{equation}
   \l \to \l' = \l ,
\end{equation}
while for $\k_i$, we have 
\begin{equation}
   \k_i \to \k_i', 
\end{equation}
such that 
\begin{equation}
   \sum_i \k'_i = \sum_i(\sfrac{1}{2} - \k_i) .
\end{equation}
It might appear that there is also a condition on
$\sum_i\k_i^2$. However, the additional fact that
$S^{-1}\circ\t_X\circ S$ is an involution, reduces this condition to
that above. 

In summary, necessary conditions for $V$ to be invariant are the
relations 
\begin{equation}
\begin{aligned}
   \t_B(\eta) &= \eta , \\
   \sum_i \k_i &= \sum_i \sfrac{1}{4} = \eta\cdot c_1 .
\end{aligned}
\label{eq:40}
\end{equation}
Since the Chern classes do not completely determine the bundle, there
may be non-invariant bundles with invariant Chern classes satisfying
these conditions. To find the exactly invariant bundles requires the
details of how the $\k_i$ transform individually. 

Since physically we
are interested mostly in the Chern classes, in this paper, for
simplicity, we will restrict ourselves to the necessary conditions
given above, without considering the details of identifying within
this class of bundles the specific ones which are exactly invariant
under $\t_X$. Therefore, we take the condition that finding
holomorphic vector bundles $V_{Z}$ over the torus-fibered threefold
$Z=X/\tau_{X}$ reduces to constructing holomorphic vector bundles $V$
over the elliptically fibered threefold $X$ that satisfy condition
\eqref{eq:40}.

The Chern classes of a vector bundle $V_{Z}$ can be determined from
the Chern classes of the $\tau_{X}$ invariant vector bundle $V$ from
which it descends as follows. Let $q:X \rightarrow Z$ be the quotient
map. Since $X$ is a double cover of $Z$, it follows that
\begin{equation}
   c_{i}(V_{Z}) = \frac{1}{2}q_{*}c_{i}(V) ,
\label{eq:42}
\end{equation}
where $q_{*}c_{i}(V)$ is the push-forward of any Chern class
$c_{i}(V)$ of $V$.


\section{Rules for Realistic Particle Physics Vacua}
\label{sec:rules}

In this section, we give the rules required to construct realistic 
particle physics vacua with $\cN=1$ supersymmetry, three families of
quarks and leptons with the standard model gauge group $\sm$. For
simplicity, we will restrict ourselves in this paper to considering
vector bundles with the structure group $SU(n)$ for $n$ odd. In the
observable sector we will choose the structure group $G=SU(5)$. The
generalization to other structure groups is relatively
straightforward. 

The first set of rules deals with the selection of the
elliptically fibered Calabi--Yau threefold $X$ which admits a
freely-acting involution $\t_X$, and the construction of bundles on
$X$ invariant under $\t_X$. As we have discussed, this allows us to
construct a torus-fibered Calabi--Yau threefold $Z=X/\t_X$, with
non-trivial fundamental group $\p_1(Z)=\ZZ_2$, and also to describe
the bundles $V_Z$ on $Z$. These rules are summaries of the conditions we have
already derived in the last two sections. If one was using this
construction to build vector bundles for each of the two $E_8$
groups in Ho\v rava-Witten theory, then this first set of constraints
is applicable to each bundle individually. The rules are 
\begin{itemize}
\item Involution conditions: Start with an elliptically fibered
   Calabi--Yau threefold $X$. If the base $B$ is smooth it must be
   either (i) a del Pezzo, (ii) a Hirzebruch, (iii) a blown-up
   Hirzebruch or (iv) an Enriques surface. To admit a freely-acting
   involution $\t_X$, we required $X$ to have two sections $\s$ and
   $\z$. In the Weierstrass model, this corresponds to an explicit
   choice of
   \begin{equation}
      g_{2} = 4(a^{2}-b) , \qquad 
      g_{3} = 4ab ,
   \label{eq:43}
   \end{equation}
   with a discriminant curve 
   \begin{equation}
      \D = 4(2a^2+b)(a^2-4b) = 0 , 
   \label{eq:44}
   \end{equation}
   The existence of $\t_X$ further required that 
   \begin{equation}
      \t_B(a) = a , \qquad
      \t_B(b) = b
   \label{abcond} 
   \end{equation}
   where $\t_B$ is the projection of $\t_X$ onto the base, and in
   addition 
   \begin{equation}
      \fixedset \cap \{\D=0\} = \emptyset ,
   \label{fscond}
   \end{equation}
   where $\fixedset$ is the set of fixed points of $\t_B$. It is
   helpful to note that this last
   condition implies that $\fixedset$ must be finite, except when the 
   base $B$ is a $dP_9$ or a K3 surface. 
\item Bundle condition: semi-stable holomorphic vector bundles over
   $X$ are given in terms of spectral data $(C,\cN)$. The
   condition that $c_1(V)=0$ implies that the spectral data can be
   written, via (\ref{eq:32}),(\ref{eq:33}),(\ref{eq:34}),
   in terms of a class $\eta$ in $B$ and coefficients $\l$ and
   $\k_i$, which satisfy
   \begin{equation}
      \l - \sfrac{1}{2} \in \ZZ , \qquad
      \k_i - \shf m \in \ZZ , 
   \label{bcond}
   \end{equation}
   with $m$ integer. Furthermore, since $C$ is an actual surface in
   the Calabi--Yau manifold, we must also have
   \begin{equation}
      \text{$\eta$ is effective}
   \end{equation}
   as a class on $B$. 
\item Bundle involution conditions: in order for $V$ to descend to a
   vector bundle $V_{Z}$ over $Z$, the class $\eta$ in $B$ and the
   coefficients $\k_i$ must satisfy the constraints 
   \begin{equation} 
   \begin{aligned} 
      \tau_{B}(\eta) &= \eta , \\
      \sum_i \k_i &= \eta\cdot c_1 .
   \end{aligned}
   \label{Vcond}
   \end{equation}
\end{itemize}

The second set of rules is directly particle physics related. In
general, in Ho\v{r}ava--Witten theory, we are free to choose two
vector bundles, $V_{Z1}$ and $V_{Z2}$, located on the ``observable''
and ``hidden'' orbifold planes respectively. In this paper, for
simplicity, we will always take $V_{Z2}$ to be the trivial
bundle. Hence, the gauge group $E_{8}$ remains unbroken on the hidden
orbifold plane.

On the observable orbifold plane, the structure group of the bundle is
$G\subset E_8$. The commutant subgroup $H$ in $E_8$ is then the
group preserved by the bundle. In this paper, we will spontaneously
break $H$ to the standard model group $\sm$ by means of a $\ZZ_2$
Wilson line on $Z$~\cite{bos,W85}. This is possible since, by construction,
we have $\p_1(Z)=\ZZ_2$. To achieve such a breaking requires us
to take 
\begin{equation}
   G = SU(5) . 
\label{eq:60}
\end{equation}
so that the unification group will be 
\begin{equation}
   H = SU(5) . 
\label{eq:61}
\end{equation}
Thus, we will assume that bundle on the observable plane $V_{Z1}$ has
structure group $G=SU(5)$.  

The first of the particle physics conditions is the requirement that
the theory have three families of quarks and leptons in the visible
sector. The number of generations associated with the vector bundle
$V_{Z1}$ over $Z$ is given by~\cite{curio} 
\begin{equation}
   N_{\text{gen}} = \sfrac{1}{2}c_{3}(V_{Z1}) . 
\label{eq:49}
\end{equation}
Requiring $N_{\text{gen}}=3$ then, using \eqref{eq:37} and
\eqref{eq:42}, leads to the following rule for the associated vector
bundle $V$ over $X$. 
\begin{itemize}
\item Three-family condition: To have three families we must require
   \begin{equation}
      \lambda \eta( \eta-nc_{1}) = 6 .
\label{3fcond}
\end{equation}
\end{itemize}

The second physical rule is associated with the anomaly cancellation
requirement~\cite{W} that 
\begin{equation}
   [W_{Z}] = c_{2}(TZ) - c_{2}(V_{Z1}) - c_{2}(V_{Z2}) ,
\label{eq:51}
\end{equation}
where $[W_{Z}]$ is the class associated with non-perturbative
five-branes in the bulk space of the theory. 
Since $V_{Z2}$ is trivial by assumption, $c_2(V_{Z2})$ vanishes and
condition \eqref{eq:51} simplifies accordingly. Using equations
\eqref{eq:29B}, \eqref{eq:42} and the fact that, by definition,
\begin{equation}
   [W_{Z}] = \hf q_{*}[W] ,
\label{eq:51A}
\end{equation}
condition \eqref{eq:51} can be pulled-back onto $X$ to give
\begin{equation}
   [W] = c_{2}(TX) - c_{2}(V) .
\label{eq:52}
\end{equation}
Inserting expressions \eqref{eq:20} and \eqref{eq:36} gives
\begin{equation}
   [W] = \s_*\o + c(F-N) + dN
\label{eq:53}
\end{equation}
where
\begin{equation}
   \o = 12c_1-\eta
\label{eq:54}
\end{equation}
and
\begin{align}
   c &= c_2 + \left(\frac{1}{24}(n^{3}-n)+11\right)c_1^2 
         - \hf\left(\l^2-\frac{1}{4}\right)
              n\eta\left(\eta-nc_1\right)
         - \sum_i\k_i^2  , 
\label{eq:55} \\
   d &= c_2 + \left(\frac{1}{24}(n^{3}-n)-1\right)c_1^2 
         - \hf\left(\l^2-\frac{1}{4}\right)
              n\eta\left(\eta-nc_1\right)
         - \sum_i\k_i^2 + \sum_i\k_i .
\label{eq:56}
\end{align}
The class $[W_{Z}]$ must represent an actual physical holomorphic
curve in the Calabi--Yau threefold $Z$ since physical five-branes are
required to wrap around it. Hence, $[W_{Z}]$ must be an effective
class, that is, $[W_{Z}]$ must be in the Mori cone of
$H_{2}(Z,\ZZ)$. It can be shown that $[W_{Z}]$ is effective in $Z$ if
and only if its pull-back $[W]$ is an effective class in the covering
threefold $X$. Therefore, we must require that $[W]$ be in the Mori
cone of $H_{2}(X,\ZZ)$. Using equation \eqref{eq:19}, this leads to
the following rule.
\begin{itemize}
\item Effectiveness condition: For $[W]$ to be an effective class, we 
   require
   \begin{equation}
      \o \text{ is effective in $B$}, \quad c \geq 0, \quad d \geq 0 .
   \label{effcond}
   \end{equation}
\end{itemize}

Next, it is possible that our bundle with structure group $G$ may
actually have a smaller structure group. If this is the case, then the
preserved subgroup of $E_{8}$ will be larger than the commutant $H$
of $G$. Berglund and Mayer~\cite{berg-mayr} have shown that, in the
context of toric $X$, this will not be the case if the vector bundle
satisfies a further constraint. (This was also discussed by
Rajesh~\cite{R}.) In the case of $SU(5)$ one requires
\begin{itemize}
\item Stability condition: Let $G=SU(5)\subset E_8$ and $G$ be the
   structure group of the vector bundle. Then the commutant $H=SU(5)$
   in $E_8$ will be the largest subgroup preserved by the bundle if 
   \begin{equation}
      \eta \geq 5 c_{1} 
   \label{Rcond}
   \end{equation}
   \end{itemize}
Geometrically, this corresponds to requiring that the spectral cover
does not split over the base $B$. 

Finally, we recall that the point of constructing $Z$ with
$\p_1(Z)=\ZZ_2$ was that we can then include a $\ZZ_2$ Wilson
line on $Z$ to break spontaneously the $H=SU(5)$ GUT group. We break 
\begin{equation}
   SU(5) \to \sm ,
\label{eq:63}
\end{equation}
by assuming that the bundle contains a non-vanishing Wilson line with
generator  
\begin{equation}
   \mathcal{G} = \left(\begin{array}{cc}
          \mathbf{1}_3 & \\ 
          & -\mathbf{1}_2 
       \end{array}\right) .
\label{Wcond}
\end{equation}
in $H=SU(5)$ 

If one follows the above rules, the vacuum will correspond to an
$\cN=1$ supersymmetric brane-world theory with, in the observable
sector, three families of quarks and leptons and the standard model gauge
group $\sm$. Armed with these rules, we now turn to the explicit
construction of phenomenologically relevant non-perturbative vacua.


\section{Three Family Models}
\label{sec:examples}

\subsection{Example 1: $B=F_2$}

In our first example, we take the base of the Calabi--Yau threefold to
be the Hirzebruch surface
\begin{equation}
   B = F_2.
\end{equation}
As discussed in the Appendix of~\cite{don2}, the Hirzebruch surfaces
are $\PP^1$ fibrations over $\PP^1$. There are two independent classes
on $F_2$, the class of the base $\cS$ and of the fiber
$\cE$. Their intersection numbers are 
\begin{equation}
   \cS\cdot\cS = -2, \qquad
   \cS\cdot\cE = 1, \qquad
   \cE\cdot\cE = 0.
\label{eq:intc2}
\end{equation}
The first and second Chern classes of $F_2$ are given by 
\begin{equation}
   c_1(F_2) = 2\cS + 4\cE, 
\label{c1F2}
\end{equation}
and
\begin{equation}
   c_2(F_2) = 4.
\label{c2F2}
\end{equation}

We now must show that there is an elliptically fibered Calabi--Yau
threefold $X$ with $F_2$ base that admits a freely-acting involution
$\t_X$ satisfying the condition given in the previous section.
The condition~\eqref{fscond} implies that the projection $\t_B$ of
$\t_X$ to the base has only a finite number of fixed points. To define
$\t_B$, we recall that there is a single type of involution on
$\PP^1$. If $(u,v)$ are homogeneous coordinates on $\PP^1$, it can be
written as $(u,v)\to(-u,v)$. This clearly has two fixed points, namely
the origin $(0,1)$ and the point at infinity $(1,0)$ in the
$u$-plane. To construct the involution $\t_B$, we combine an
involution on the base $\cS=\PP^1$ with one on the fiber
$\cE=\PP^1$. We will not give details here, except to note that one
finds that ${\cal{F}}_{\tau_{B}}$ contains four fixed points, coming
from the fixed points of the $\PP^1$ fibers above the fixed points of
the involution on the base. We should also point out that, in general,
it is not possible to find involutions with a finite number of fixed
points for all $F_r$. 

To ensure that we can construct a freely-acting involution $\t_X$ from
$\t_B$, we further need to show~\eqref{fscond} that the discriminant
curve can be chosen so as not to pass through these fixed points. We
recall that the discriminant curve is given by 
\begin{equation}
   4\left(2a^2 + b\right)^2(a^2 - 4b) = 0, 
\end{equation}
and that the parameters $a$ and $b$ are sections of $K_B^{-2}$ and
$K_B^{-4}$ respectively, where $K_B$ is the canonical bundle of the
base. In order to lift $\t_B$ to an involution of $X$, we also 
require~\eqref{abcond} that
\begin{equation}
   \t_B(a) = a, \qquad
   \t_B(b) = b.
\label{eq:hi} 
\end{equation}
This restricts the allowed sections $a$ and $b$ and, consequently, the
form of $\Delta_{1}$ and $\Delta_{2}$. The question is then whether,
within the class of allowed sections $a$ and $b$, there are examples
where the corresponding discriminant curves avoid the fixed
points. One can easily show that such sections $a$ and $b$ do exist. 

To satisfy the conditions~\eqref{Vcond} on the invariance of $V$ under
the involution, we need to find the classes $\eta$ in $F_{2}$ that are
invariant under $\tau_{B}$. We find that the involution preserves
both $\cS$ and $\cE$ separately, so that  
\begin{equation}
   \t_B(\cS) = \cS, \qquad
   \t_B(\cE) = \cE.
\label{eq:hello}
\end{equation}
Since any class $\eta$ is a linear combination of $\cS$ and $\cE$, we
see that, in fact, an arbitrary $\eta$ satisfies $\tau_{B}(\eta)=\eta$.

We can now search for $\eta$, $\l$ and $\k_i$ satisfying the three
family~\eqref{3fcond}, effectiveness~\eqref{effcond} and
stability~\eqref{Rcond} conditions given above. We find that there are two
classes of solutions 
\begin{equation}
\begin{aligned}
   \text{solution 1:} & \quad 
       \eta = 14\mathcal{S} + 22\mathcal{E}, \quad 
       \l = \sfrac{3}{2}, \\
       {}& \sum_i\k_i = \eta\cdot c_1 = 44, \quad
       \sum_i \k_i^2 \leq 60 , \\
   \text{solution 2:} & \quad
       \eta = 24\mathcal{S} + 30\mathcal{E}, \quad 
       \l = -\sfrac{1}{2}, \\
       {}& \sum_i\k_i = \eta\cdot c_1 = 60, \quad
       \sum_i \k_i^2 \leq 76 .  
\end{aligned}
\label{solF2}
\end{equation}

First note that the coefficients $\l$ satisfy the bundle constraint
\eqref{bcond}. Furthermore, one can find many examples of $\k_i$ with
$i=1,\dots,4\eta\cdot c_1$, satisfying the bundle
constraint~\eqref{bcond}, the given conditions on $\sum_i\k_i^2$ and
the invariance condition $\sum_i\k_i=\eta\cdot c_1$. 

Using $n=5$, \eqref{c1F2}, \eqref{solF2} and the intersection
relations \eqref{eq:intc2}, one can easily verify that both
solutions satisfy the three-family condition \eqref{3fcond}. 

Next, from \eqref{eq:53}, \eqref{eq:54}, \eqref{eq:55}
and \eqref{eq:56}, as well as $n=5$, \eqref{c1F2}, \eqref{c2F2},
\eqref{solF2} and the intersection relations \eqref{eq:intc2}, we can
calculate the five-brane curves $W$ associated with each of the
solutions. We find that 
\begin{equation}
\begin{aligned}
   \text{solution 1:} \quad & 
      [W] = \s_*\left(10\cS+26\cE\right)
             + \left(112-k\right)\left(F-N\right) 
             + \left(60-k\right) N, \\   
   \text{solution 2:} \quad & 
      [W] = \s_*\left(18\cE\right)
             + \left(132-k\right)\left(F-N\right) 
             + \left(76-k\right) N, 
\end{aligned}
\label{eq:branes}
\end{equation}
where
\begin{equation}
   k = \sum_i \k_i^2
\end{equation}
It follows that the base components for $[W]$ are given by
\begin{equation}
\begin{aligned}
   \text{solution 1:} \quad & \o = 10\cS + 26\cE, \\
   \text{solution 2:} \quad & \o = 18\cE,
\end{aligned}
\end{equation}
which are both effective. Furthermore, we note that for each
five-brane curve the $c$ and $d$ coefficients of classes $F-N$ and $N$
respectively are non-negative integers (given the constraints 
(\ref{solF2}) on
$k$). Hence, effectiveness condition \eqref{effcond} is satisfied. 

Finally, note that the stability condition requires 
$\eta\geq 5c_{1}$. In both of the above solutions  
\begin{equation}
   \eta > 10\mathcal{S} + 20\mathcal{E} = 5c_1 ,
\end{equation}
so that the condition is satisfied.

We conclude that, over a Hirzebruch base $B=F_{2}$, one can construct
torus-fibered Calabi--Yau threefolds, $Z$, without section and with
non-trivial first homotopy group $\pi_{1}(Z)=\ZZ_{2}$. Assuming a
trivial gauge vacuum on the hidden brane, we have shown that we expect
these threefolds to admit two classes of semi-stable holomorphic
vector bundles $V_{Z}$, \eqref{solF2}, associated with an 
$\cN=1$
supersymmetric theory. These vacua have three families of chiral quarks and
leptons and GUT group $H=SU(5)$ on the observable brane-world. Since
$\pi_{1}(Z)=\ZZ_{2}$, Wilson lines break this GUT group 
\begin{equation}
   SU(5) \rightarrow SU(3)_{C} \times SU(2)_{L} \times U(1)_{Y} ,
\end{equation}
to the standard model gauge group. Anomaly cancellation and
supersymmetry require the existence of non-perturbative five-branes in
the extra dimension of the bulk space. These five-branes are wrapped
on holomorphic curves in $Z$ whose homology classes~\eqref{eq:branes}
are exactly calculable.

\subsection{Example 2: $B=dP_{3}$}

Let us now choose the base of the Calabi--Yau threefold $X$ to be 
\begin{equation}
   B = dP_{3} .
\label{eq:65}
\end{equation}
Recall that $dP_{3}$ can be thought of as the projective plane
$\PP^{2}$ blown-up at three points. In terms of the homogeneous
coordinates $(u,v,w)$ on $\PP^{2}$, the points we blow up may be taken
to be
\begin{equation}
   (1,0,0), \quad (0,1,0), \quad (0,0,1) .
\label{eq:65A}
\end{equation}
The blownup surface $B$ 
can be embedded in $\PP^6$ with 
homogeneous coordinates $(z_0,\dots,z_6)$ given by 
\begin{equation}
   (z_0,\dots,z_6) = (uvw,u^2v,w^2v,v^2w,u^2w,w^2u,v^2u).
\end{equation}
The blown-up points~\eqref{eq:65A} then correspond to projective
lines
in $\PP^6$. For instance, considering the limit as $v$ and $w$ go to
zero, we see that the point $(u,v,w)=(1,0,0)$ maps to the whole
$\PP^1$ given by $(z_0,\dots,z_6)=(0,v,0,0,w,0,0)$ in $\PP^6$. 

The properties of del Pezzo surfaces were reviewed in the Appendix
of~\cite{don2}. For $dP_{3}$, one finds a basis for $H_{2}(dP_{3},\ZZ)$
composed entirely of effective classes given by $l$ (the pullback
of
the hyperplane class of $\PP^{2}$) and three exceptional divisors
$E_{i}$ for $i=1,2,3$ corresponding to the three blown-up
points. Their intersection numbers are 
\begin{equation}
   l\cdot l = 1 , \quad 
   l\cdot E_{i} = 0 , \quad 
   E_{i}\cdot E_{j} = -\delta_{ij} .
\label{eq:66}
\end{equation}
In addition, there are three other dependent exceptional classes
\begin{equation}
   l-E_{2}-E_{3} , \quad 
   l-E_{3}-E_{1} , \quad 
   l-E_{1}-E_{2} ,
\label{eq:67}
\end{equation}
corresponding to the lines in $\PP^{2}$ passing through pairs of the
blown-up points. These lines are given by 
\begin{equation}
   u=0 , \quad 
   v=0 , \quad 
   w=0 
\label{eq:68}
\end{equation}
respectively. The first and second Chern classes for $dP_{3}$ are
\begin{equation}
   c_{1}(dP_{3}) = 3l - E_{1} - E_{2} - E_{3} ,
\label{eq:69}
\end{equation}
and
\begin{equation}
   c_{2}(dP_{3}) = 6 .
\label{eq:70}
\end{equation}

First, we must again show that the Calabi--Yau threefold $X$ based on
$B=dP_3$ admits a freely-acting involution. The
requirement~\eqref{fscond} implies that the projection $\t_B$ of
$\t_X$ to the base must have a finite number of fixed points. We
start, therefore, by showing that such a $\t_B$ exists. In terms of the
$\PP^{2}$ coordinates $(u,v,w)$ define $\t_B$ by  
\begin{equation}
   \t_B: (u,v,w) \rightarrow (u^{-1},v^{-1},w^{-1}) .
\label{eq:71}
\end{equation}
In the ambiant space $\PP^6$ space this is simply the map exchanging $z_1$
and $z_2$, $z_3$ and $z_4$, and $z_5$ and $z_6$. Thus it is clearly an
involution of the $dP_{3}$. Furthermore, it follows from \eqref{eq:71}
that $\tau_{B}$ has four isolated fixed points
\begin{equation}
   \mathcal{F}_{\t_B} =\{(1,\pm1, \pm1)\} .
\label{eq:72}
\end{equation}

To ensure that we can construct a freely-acting involution $\t_X$ from
$\t_B$, we need to show that the discriminant curve can be chosen so
as not to intersect these fixed points. We recall that the two
components of the discriminant curve are given by 
\begin{equation}
   \D_1 = a^2 - 4b = 0, \qquad
   \D_2 = 4\left(2a^2 + b\right) = 0,
\end{equation}
and that the parameters $a$ and $b$ are sections of $K_B^{-2}$ and
$K_B^{-4}$ respectively, where $K_B$ is the canonical bundle of the
base. The first Chern class of the anti-canonical bundle $K_B^{-1}$ is
$c_1(B)$. Thus from~\eqref{eq:69}, we see that as classes
\begin{equation}
   [a] = 2\left(3l - E_1 - E_2 - E_3\right), \qquad
   [b] = 4\left(3l - E_1 - E_2 - E_3\right). 
\end{equation}
In terms of the blown-up $\PP^2$, $a$ and $b$ each correspond to
curves. The function $a$ is a degree six homogeneous polynomial
in $(u,v,w)$, describing a curve which passes twice through each of
the blown-up points. The function $b$ is a degree twelve polynomial and
describes a curve passing four times through each blow-up. These
conditions restrict the form of the polynomials. For instance, all the
terms involving fifth- or sixth-order powers of $u$, $v$ or $w$ are
excluded in $a$. In addition, in order to lift $\t_B$ to an involution
of $X$, recall that we required 
\begin{equation}
   \t_B(a) = a, \qquad
   \t_B(b) = b. 
\end{equation}
This means that the polynomials are further restricted to be
homogeneous functions of $(u,v,w)$ which are invariant under the
involution~\eqref{eq:71}. It is a simple process to identify all such
polynomials. The discriminant curves are then given by the vanishing
of the twelfth-order polynomials $\D_1$ and $\D_2$. The question is
then whether, within the class of allowed $a$ and $b$ polynomials,
there are examples where the corresponding discriminant curves avoid
the fixed points. In fact, from the form of $a$ and $b$, one can show
that there is enough freedom in choosing the polynomials, so that any
generic choice leads to discriminant curves which do not intersect any
of the fixed points.

In order to have invariant bundles on $X$, by the
condition~\eqref{Vcond}, we need to find classes $\eta$ in $dP_3$ that
satisfy $\tau_{B}(\eta)=\eta$. Using \eqref{eq:65A}, \eqref{eq:68} and
the definition of $\tau_{B}$, we find that 
\begin{equation}
\begin{aligned}
   \t_B(E_1) &= l - E_2 - E_3 , \\
   \t_B(E_2) &= l - E_3 - E_1 , \\
   \t_B(E_3) &= l - E_1 - E_2 .
\end{aligned}
\label{eq:73}
\end{equation}
Since the involution must preserve the intersection numbers and map
effective curves to effective curves, this implies that 
\begin{equation}
   \t_B(l) = 2l - E_1 - E_2 - E_3 .
\label{eq:74}
\end{equation}
From these expressions, we see that there are three independent 
curves:
\begin{equation}
\begin{aligned}
   a_1 &= l + E_1 - E_2 - E_3 , \\
   a_2 &= l - E_1 + E_2 - E_3 , \\
   a_3 &= l - E_1 - E_2 + E_3 ,
\end{aligned}
\label{eq:75}
\end{equation}
satisfying
\begin{equation}
\tau_{B}(a_i) = a_i
\label{eq:76}
\end{equation}
for $i=1,2,3$, and one perpendicular class
\begin{equation}
   p= l - E_1 - E_2 - E_3 ,
\label{eq:77}
\end{equation}
satisfying
\begin{equation}
   \tau_{B}(p)=- p .
\label{eq:78}
\end{equation}
The three classes $a_{i}$ for $i=1,2,3$ generate all other $\tau_{B}$
invariant curves. Note that each class $a_{i}$ can be written as a sum
of effective classes with non-negative integer coefficients. For
example, $a_{1}=E_{1}+(l-E_{2}-E_{3})$. Hence, each $a_{i}$ is an
effective class. It follows that the condition $\t_B(\eta)=\eta$ can be
solved simply by demanding that $\eta$ be written in terms of
invariant classes only. That is
\begin{equation}
   \eta = m_1 a_1 + m_2 a_2 + m_3 a_3 ,
\label{eq:79}
\end{equation}
where the $m_{i}$ are constant coefficients. In addition, note that
the first Chern class of $dP_3$ given in \eqref{eq:69} can be written
as 
\begin{equation}
   c_1 = a_1 + a_2 + a_3
\label{eq:80}
\end{equation}
and, hence, is invariant under $\t_B$, as it must be. 

We can now search for $\eta$, $\l$ and $\k_{i}$ satisfying the three
family~\eqref{3fcond}, effectiveness~\eqref{effcond} and
stability~\eqref{Rcond} conditions given above. We find that there are
three classes of solutions  
\begin{equation}
\begin{aligned}
   \text{solution 3:} & \quad 
       \eta = 17l - 3E_1 - 7E_2 - 7E_3, \quad
       \l = \sfrac{1}{2}, \\
       {}& \sum_i\k_i = \eta\cdot c_1 = 34 , \quad
       \sum_i\k_i^2 \leq 64 , \\
   \text{solution 4:} & \quad
       \eta = 18l - 2E_1 - 8E_2 - 8E_3, \quad
       \l = \sfrac{1}{2}, \\
       {}&\sum_i\k_i = \eta\cdot c_1 = 36 , \quad
       \sum_i\k_i^2 \leq 66 , \\
   \text{solution 5:} & \quad
       \eta = 21l + E_1 - 11E_2 - 11E_3, \quad
       \l = -\sfrac{1}{2}, \\
       {}&\sum_i\k_i = \eta\cdot c_1 = 42 , \quad
       \sum_i\k_i^2 \leq 72 , \\
\end{aligned}
\label{eq:81}
\end{equation}

First note that the coefficients $\l$ satisfy the bundle constraint
\eqref{bcond}. Furthermore, one can find many examples of $\k_i$ with
$i=1,\dots,4\eta\cdot c_1$, satisfying the bundle
constraint~\eqref{bcond}, the given conditions on $\sum_i\k_i^2$ and
the invariance condition $\sum_i\k_i=\eta\cdot c_1$. 

Second, we see that each curve $\eta$ can be expressed as
\begin{equation}
\begin{aligned}
   \text{solution 3:} \quad & \eta = 7a_1 + 5a_2 + 5a_3 , \\
   \text{solution 4:} \quad & \eta = 8a_1 + 5a_2 + 5a_3 , \\
   \text{solution 5:} \quad & \eta = 11a_1 + 5a_2 + 5a_3 , 
\end{aligned}
\label{eq:82}
\end{equation}
and, therefore, each $\eta$ is invariant under $\t_B$. Using $n=5$,
\eqref{eq:69}, \eqref{eq:81} and the intersection relations
\eqref{eq:66}, one can easily verify that all three solutions satisfy
the three-family condition \eqref{3fcond}. Now, from \eqref{eq:53},
\eqref{eq:54}, \eqref{eq:55} and \eqref{eq:56}, as well as $n=5$,
\eqref{eq:69}, \eqref{eq:70}, \eqref{eq:81} and the intersection
relations \eqref{eq:66}, we can calculate the five-brane curves $W$
associated with each of the three solutions. We find that 
\begin{equation}
\begin{aligned}
   \text{solution 3:} \quad & 
      [W] = \s_*\left(5a_1+7a_2+7a_3\right)
            + \left(102-k\right)\left(F-N\right) 
            + \left(64-k\right)N , \\   
   \text{solution 4:} \quad & 
      [W] = \s_*\left(4a_1+7a_2+7a_3\right)
            + \left(102-k\right)\left(F-N\right) 
            + \left(66-k\right)N , \\
   \text{solution 5:} \quad & 
      [W] = \s_*\left(a_1+7a_2+7a_3\right)
            + \left(102-k\right)\left(F-N\right) 
            + \left(72-k\right)N ,
\end{aligned}
\label{eq:83}
\end{equation}
where, as before, 
\begin{equation}
   k = \sum_i \k_i^2 .
\end{equation}
It follows that the base components for $[W]$ are given by
\begin{equation}
\begin{aligned}
   \text{solution 3:} \quad & \o = 5a_1 + 7a_2 + 7a_3 , \\
   \text{solution 4:} \quad & \o = 4a_1 + 7a_2 + 7a_3 , \\
   \text{solution 5:} \quad & \o = a_1 + 7a_2 + 7a_3 . \\
\end{aligned}
\label{eq:84}
\end{equation}
Since the $a_i$ are effective, so is each of these $\o$
classes. Furthermore, we note, given the conditions on $k$, that for
each five-brane curve the $c$ and $d$ coefficients of the classes $F-N$
and $N$ respectively are all non-negative integers. Hence, the
effectiveness condition \eqref{effcond} is satisfied. 

Finally, note that the stability condition requires $\eta\geq 5c_{1}$. In
all of the above solutions   
\begin{equation}
   \eta > 5a_1+5a_2+5a_3 = c_1
\label{eq:85} 
\end{equation}
so the condition is satisfied.

We conclude that, over a del Pezzo base $B=dP_3$, one can also
construct torus-fibered Calabi--Yau threefolds, $Z$, without section
and with non-trivial first homotopy group
$\pi_{1}(Z)=\ZZ_{2}$. Assuming a trivial gauge vacuum on the hidden
brane, we have shown that we expect these threefolds to admit
three families~\eqref{eq:18} of semi-stable holomorphic vector bundles
$V_{Z}$, associated with an $\cN=1$ supersymmetric theory
with three families of chiral quarks and leptons and GUT group
$H=SU(5)$ on the observable brane-world. Since $\pi_{1}(Z)=\ZZ_{2}$,
Wilson lines break this GUT group 
\begin{equation}
   SU(5) \rightarrow SU(3)_{C} \times SU(2)_{L} \times U(1)_{Y} ,
\end{equation}
to the standard model gauge group. Anomaly cancellation and
supersymmetry require the existence of non-perturbative five-branes in
the extra dimension of the bulk space. These five-branes are wrapped
on holomorphic curves in $Z$ whose homology classes, \eqref{eq:83},
are exactly calculable.


\section{Conclusions}
\label{sec:concl}

In view of the technical nature of this paper, it is useful to
conclude by focusing on the structure and physical properties of
these ``standard model'' supersymmetric heterotic M-theory vacua. 

The heterotic M-theory vacua discussed in this paper 
are constructed by compactifying Ho\v rava--Witten theory on a
smooth Calabi--Yau threefold, $Z$, one of a specific class with the
property that
\begin{equation}
   \p_1 \left(Z\right) = \ZZ_2 .
\label{Bz2}
\end{equation}
The threefolds discussed here were constructed from elliptically
fibered Calabi--Yau threefolds, $X$, with two sections, by modding out
a free involution that exchanges the two sections. The resulting
Calabi--Yau spaces, $Z$, are torus-fibered but not elliptically
fibered since they admit no global sections. For instance in
section~\ref{sec:examples}, example 2 we constructed $Z$ from an
elliptically fibered threefold $X$ with base $B=dP_3$, by modding out
the involution which, restricted to the base, acts as
\begin{equation}
   (u,v,w) \to (u^{-1},v^{-1},w^{-1}) ,
\label{BdP3z2}
\end{equation}
where $u$, $v$ and $w$ are complex coordinates on the $dP_3$. 

The vacua of the effective low-energy theory consist of a large
five-dimensional bulk space bounded on each end of the fifth dimension
by a BPS domain wall with $\cN=1$ supersymmetry on its world volume. A
priori, these walls carry the $E_8$ gauge supermultiplets of the
heterotic M-theory. We call one of the walls the ``observable''
sector and the other the ``hidden'' sector of the theory. 

First consider the observable wall. In this paper, we constructed a
class  of smooth semi-stable holomorphic
vector bundles over $X$ with the structure group
\begin{equation}
   G = SU(5) .
\label{BG}
\end{equation}
whose Chern classes are invariant under the involution $\tau_X$. The
involution acts on this class of bundles; the invariant ones are
precisely the bundles which come from $Z$. Their detailed
construction is not required here since it is only the Chern classes that
enter the physical discussion. The description of the invariant bundles will
be given in a forthcoming paper \cite{mathpaper}.

Each such bundle corresponds to some $SU(5)$ instanton configuration
on $Z$, which preserves $\cN=1$ supersymmetry. However, it
spontaneously breaks $E_8$ to the unification group $H=SU(5)$, which
is the commutant of $G=SU(5)$ in $E_8$. Thus the gauge group on the
observable wall is only $H=SU(5)$. 

These bundles were constructed by first forming semi-stable
holomorphic $SU(5)$ bundles over $X$ via the spectral cover
construction, and then restricting to bundles that were invariant under
the involution. Such bundles ``descend'' to $Z$. For instance, in
example 3 in section~\ref{sec:examples}, the spectral data, specifying
the bundle over $X$, were the class of a curve
\begin{equation}
  \eta = 17l - 3E_1 - 7E_2 - 7E_3
\label{Beta}
\end{equation}
in $B=dP_3$, together with the coefficients
\begin{equation}
   \l = \shf
\label{Blm}
\end{equation}
and $\k_i$ satisfying
\begin{equation}
   \sum_i \k_i = 34 , \qquad 
   \sum_i \k_i^2 \leq 64 .
\label{Bkappa}
\end{equation}
This data specifies a class of $G=SU(5)$ 
semi-stable holomorphic bundle over $X$
whose Chern classes are invariant under the
involution $\t_X$. The involution $\t_X$ therefore acts on the moduli 
space of these bundles. The fixed points of this action correspond to 
bundles which descend to a $G=SU(5)$ semi-stable holomorphic bundle on
$Z$. 

In addition, with the given choices of $\l$
and $\eta$, the associated vector bundle $V_Z$ over $Z$ has third Chern
class $c_3(V_Z)=6$. This implies that the number of generations of
quarks and leptons in the observable sector is 
\begin{equation}
   N_{\text{gen}} = 3 .
\label{Bgen}
\end{equation}

Finally, since $\p_1(Z)=\ZZ_2$, the $G=SU(5)$ gauge bundles can be
``enlarged'' by $\ZZ_2$ Wilson lines. These spontaneously break the
unification group $SU(5)$ as 
\begin{equation}
   SU(5) \to SU(3)_C \times SU(2)_L \times U(1)_Y 
\label{Bbreak}
\end{equation}
on the observable domain wall.

Now consider the hidden wall. In this paper, for simplicity, we assume
that we take a trivial vector bundle in this sector. Hence, $E_8$ is
the unbroken gauge group on the hidden sector domain wall. 

The vector bundles of the observable and hidden sectors are linked to
the structure of $Z$ by the requirement of anomaly
cancellation. Specifically, this relates the second Chern classes of
the vector bundles to the second Chern class of the tangent bundle of
$Z$. We find that, for three families in the observable sector,
anomaly cancellation implies the existence of five-branes in the bulk
space wrapped around a holomorphic curve in $Z$. The homology
class of this curve, $W$, can be explicitly computed given $Z$ and the
two vector bundles. For instance, for example 2 of
section~\ref{sec:examples}, we showed that 
\begin{equation}
   [W] = \s_*\left(19l - 9E_1 - 5E_2 - 5E_3\right)
            + \left(102-k\right)\left(F-N\right) 
            + \left(64-k\right)N
\end{equation}
where $k=\sum_i\k_i^2$. The generic structure of these ``standard''
model heterotic M-theory vacua is shown pictorially in
Figure~\ref{fig:braneworld}.


\subsection*{Acknowledgments}

R.~Donagi is supported in part by an NSF grant DMS-9802456 as well as a
UPenn Research Foundation Grant. He was partially supported by a grant
from the Emmy Noether Institute and the Minerva foundation of Germany, as
well as by the Hebrew University of Jerusalem and RIMS, Kyoto during
visits to those institutiuons. 
B.~A.~Ovrut is supported in part by a Senior Alexander von Humboldt
Award, by the DOE under contract No. DE-AC02-76-ER-03071 and by a
University of Pennsylvania Research Foundation Grant. 
T.~Pantev is supported in part by an NSF grant DMS-9800790 and by an
Alfred P. Sloan Research Fellowship. 
D.~Waldram was supported in part at Princeton University by the DOE
under contract No. DE-FG02-91ER40671 and also wishes to acknowledge
the hospitality of the Aspen Center for Physics where part of this
work was done.




\begin{thebibliography}{99}

\bibitem{HW1}
   P.~Horava and E.~Witten,
   ``Heterotic and Type I String Dynamics from Eleven Dimensions,''
   Nucl.\ Phys.\  {\bf B460}, 506 (1996)
   [hep-th/9510209].
\bibitem{HW2}
   P.~Horava and E.~Witten,
   ``Eleven-Dimensional Supergravity on a Manifold with Boundary,''
   Nucl.\ Phys.\  {\bf B475}, 94 (1996)
   [hep-th/9603142].
\bibitem{W}
   E.~Witten,
   ``Strong Coupling Expansion Of Calabi-Yau Compactification,''
   Nucl.\ Phys.\  {\bf B471}, 135 (1996)
   [hep-th/9602070].
\bibitem{scales}
   T.~Banks and M.~Dine,
   ``Couplings and Scales in Strongly Coupled Heterotic String Theory,''
   Nucl.\ Phys.\  {\bf B479}, 173 (1996)
   [hep-th/9605136];
   \\
   I.~Antoniadis and M.~Quiros,
   ``Large radii and string unification,''
   Phys.\ Lett.\  {\bf B392}, 61 (1997)
   [hep-th/9609209];
   \\
   K.~Benakli,
   ``Phenomenology of low quantum gravity scale models,''
   Phys.\ Rev.\  {\bf D60}, 104002 (1999)
   [hep-ph/9809582].
\bibitem{dim} 
   N.~Arkani-Hamed, S.~Dimopoulos and G.~Dvali,
   ``The hierarchy problem and new dimensions at a millimeter,''
   Phys.\ Lett.\  {\bf B429}, 263 (1998)
   [hep-ph/9803315];
   \\
   I.~Antoniadis, N.~Arkani-Hamed, S.~Dimopoulos and G.~Dvali,
   ``New dimensions at a millimeter to a Fermi and superstrings at a TeV,''
   Phys.\ Lett.\  {\bf B436}, 257 (1998)
   [hep-ph/9804398].
\bibitem{RS}
   L.~Randall and R.~Sundrum,
   ``A large mass hierarchy from a small extra dimension,''
   Phys.\ Rev.\ Lett.\  {\bf 83}, 3370 (1999)
   [hep-ph/9905221];
   \\
   L.~Randall and R.~Sundrum,
   ``An alternative to compactification,''
   Phys.\ Rev.\ Lett.\  {\bf 83}, 4690 (1999)
   hep-th/9906064.
\bibitem{nse}
   A.~Lukas, B.~A.~Ovrut and D.~Waldram,
   ``Non-standard embedding and five-branes in heterotic M-theory,''
   Phys.\ Rev.\  {\bf D59}, 106005 (1999)
   [hep-th/9808101].
\bibitem{lpt}
   Z.~Lalak, S.~Pokorski and S.~Thomas,
   ``Beyond the standard embedding in M-theory on S(1)/Z(2),''
   Nucl.\ Phys.\  {\bf B549}, 63 (1999)
   [hep-ph/9807503].
\bibitem{stei}
   S.~Stieberger,
   ``(0,2) heterotic gauge couplings and their M-theory origin,''
   Nucl.\ Phys.\  {\bf B541}, 109 (1999)
   [hep-th/9807124].
\bibitem{losw1}
   A.~Lukas, B.~A.~Ovrut, K.~S.~Stelle and D.~Waldram,
   ``The universe as a domain wall,''
   Phys.\ Rev.\  {\bf D59}, 086001 (1999)
   [hep-th/9803235].
\bibitem{losw2}
   A.~Lukas, B.~A.~Ovrut, K.~S.~Stelle and D.~Waldram,
   ``Heterotic M-theory in five dimensions,''
   Nucl.\ Phys.\  {\bf B552}, 246 (1999)
   [hep-th/9806051].
\bibitem{Don} 
   S.~Donaldson, 
   Proc. London Math. Soc. {\bf 3} 1 (1985).
\bibitem{UhYau} 
   K.~Uhlenbeck and S.-T.~Yau, 
   Comm. Pure App. Math. {\bf 39} 257 (1986), {\bf 42} 703 (1986).
\bibitem{dg}
   J.~Distler and B.~Greene,
   ``Aspects Of (2,0) String Compactifications,''
   Nucl.\ Phys.\  {\bf B304}, 1 (1988).
\bibitem{kachru}
   S.~Kachru,
   ``Some three generation (0,2) Calabi-Yau models,''
   Phys.\ Lett.\  {\bf B349}, 76 (1995)
   [hep-th/9501131].
\bibitem{FMW} 
   R.~Friedman, J.~Morgan and E.~Witten,
   ``Vector bundles and F theory,''
   Commun.\ Math.\ Phys.\  {\bf 187}, 679 (1997)
   [hep-th/9701162].
\bibitem{D}
   R.~Donagi, 
   Asian. J. Math. {\bf 1}, 214 (1997).
\bibitem{BJPS} 
   M.~Bershadsky, A.~Johansen, T.~Pantev and V.~Sadov,
   ``On four-dimensional compactifications of F-theory,''
   Nucl.\ Phys.\  {\bf B505}, 165 (1997)
   [hep-th/9701165].
\bibitem{don1}
   R.~Donagi, A.~Lukas, B.~A.~Ovrut and D.~Waldram,
   ``Non-perturbative vacua and particle physics in M-theory,''
   JHEP {\bf 9905}, 018 (1999)
   [hep-th/9811168].
\bibitem{don2}
   R.~Donagi, A.~Lukas, B.~A.~Ovrut and D.~Waldram,
   ``Holomorphic vector bundles and non-perturbative vacua in M-theory,''
   JHEP {\bf 9906}, 034 (1999)
   [hep-th/9901009].
\bibitem{curio}
   G.~Curio,
   ``Chiral matter and transitions in heterotic string models,''
   Phys.\ Lett.\  {\bf B435}, 39 (1998)
   [hep-th/9803224].
\bibitem{ba} 
   B.~Andreas,
   ``On vector bundles and chiral matter in N = 1 heterotic
      compactifications,'' 
   JHEP {\bf 9901}, 011 (1999)
   [hep-th/9802202].
\bibitem{don3} 
   R.~Donagi, B.~A.~Ovrut and D.~Waldram,
   ``Moduli spaces of fivebranes on elliptic Calabi-Yau threefolds,''
   JHEP {\bf 9911}, 030 (1999)
   [hep-th/9904054].
\bibitem{bos} 
   J.~D.~Breit, B.~A.~Ovrut and G.~C.~Segre,
   ``E(6) Symmetry Breaking In The Superstring Theory,''
   Phys.\ Lett.\  {\bf B158}, 33 (1985);
\bibitem{W85}
   E.~Witten,
   ``Symmetry Breaking Patterns In Superstring Models,''
   Nucl.\ Phys.\  {\bf B258}, 75 (1985).
\bibitem{mathpaper}
   R.~Donagi, B.~A.~Ovrut, T.~Pantev and D.~Waldram,
   in preparation.
\bibitem{ACK}
   B.~Andreas, G.~Curio and A.~Klemm,
   ``Towards the standard model spectrum from elliptic Calabi-Yau,''
   hep-th/9903052.
\bibitem{MV}
   A.~Grassi, 
   Internat. J. Math. {\bf 4}, 203 (1993); \\
   D.~R.~Morrison and C.~Vafa,
   ``Compactifications of F-Theory on Calabi--Yau Threefolds -- II,''
   Nucl.\ Phys.\  {\bf B476}, 437 (1996)
   [hep-th/9603161].
\bibitem{thomas}
   R.~P.~Thomas,
   ``Examples of bundles on Calabi-Yau 3-folds for string theory
      compactifications,'' 
   [math.AG/9912179].
\bibitem{Dcharge}
   M.~Li,
   ``Boundary States of D-Branes and Dy-Strings,''
   Nucl.\ Phys.\  {\bf B460}, 351 (1996)
   [hep-th/9510161]; 
   \\
   M.~B.~Green, J.~A.~Harvey and G.~Moore,
   ``I-brane inflow and anomalous couplings on D-branes,''
   Class.\ Quant.\ Grav.\  {\bf 14}, 47 (1997)
   [hep-th/9605033].
\bibitem{AD}
   P.~S.~Aspinwall and R.~Y.~Donagi,
   ``The heterotic string, the tangent bundle, and derived categories,''
   Adv.\ Theor.\ Math.\ Phys.\  {\bf 2}, 1041 (1998)
   [hep-th/9806094].
\bibitem{berg-mayr}
   P.~Berglund and P.~Mayr,
   ``Heterotic string/F-theory duality from mirror symmetry,''
   Adv.\ Theor.\ Math.\ Phys.\  {\bf 2}, 1307 (1999)
   [hep-th/9811217].
   \\
   P.~Berglund and P.~Mayr,
   ``Stability of vector bundles from F-theory,''
   JHEP {\bf 9912}, 009 (1999)
   [hep-th/9904114].
\bibitem{R}
   G.~Rajesh,
   ``Toric geometry and F-theory/heterotic duality in four dimensions,''
   JHEP {\bf 9812}, 018 (1998)
   [hep-th/9811240].

\end{thebibliography}
\end{document}